\documentclass[aip,reprint]{revtex4-2}
\usepackage[a4paper, margin=1in]{geometry}
\usepackage{graphicx} 
\usepackage[usenames,dvipsnames]{xcolor}
\usepackage{amsmath}
\usepackage{caption}
\usepackage{enumitem}
\usepackage{amssymb}
\usepackage{wrapfig}
\usepackage{color}
\usepackage{booktabs}
\usepackage{array} 
\usepackage{scrextend}
\usepackage{physics}
\usepackage{mathrsfs}
\usepackage{float}
\usepackage{bbm}
\usepackage{subcaption}
\usepackage{caption}
\usepackage{comment}
\usepackage{epstopdf}
\usepackage{amsmath} 
\usepackage{siunitx}
\usepackage{subfig}
\usepackage{listings}
\usepackage[american]{circuitikz}
\usepackage{tikz}
\usetikzlibrary{shapes.arrows}
\usetikzlibrary{arrows.meta}
\makeatletter
\def\TikzBipolePath#1#2{\pgf@circ@bipole@path{#1}{#2}}

\makeatother
\newlength{\ResUp}
\newlength{\ResDown}
\newlength{\ResLeft}
\newlength{\ResRight}
\ctikzset{bipoles/josephsonjunction/height/.initial=.30}   
\ctikzset{bipoles/josephsonjunction/width/.initial=.30}    
\pgfcircdeclarebipole{}                                    
{\ctikzvalof{bipoles/josephsonjunction/height}}
{josephsonjunction}                                        
{\ctikzvalof{bipoles/josephsonjunction/height}}
{\ctikzvalof{bipoles/josephsonjunction/width}}
{                                                          
 \pgfextracty{\ResUp}{\northeast}                         
 \pgfextracty{\ResDown}{\southwest}
 \pgfextractx{\ResLeft}{\southwest}
 \pgfextractx{\ResRight}{\northeast}
 \pgfsetlinewidth{3\pgfstartlinewidth}
 \pgfmoveto{\pgfpoint{\ResLeft}{\ResDown}}
 \pgflineto{\pgfpoint{\ResRight}{\ResUp}}
 \pgfmoveto{\pgfpoint{\ResRight}{\ResDown}}
 \pgflineto{\pgfpoint{\ResLeft}{\ResUp}}
 \pgfusepath{draw}
 \pgfsetlinewidth{\pgfstartlinewidth}
 \pgfmoveto{\pgfpoint{\ResLeft}{0}}
 \pgflineto{\pgfpoint{\ResRight}{0}}
 \pgfusepath{draw}
}
\def\circlepath#1{\TikzBipolePath{josephsonjunction}{#1}}
\tikzset{josephsonjunction/.style = {\circuitikzbasekey, /tikz/to path=\circlepath, l=#1}}

\makeatletter
\def\@email#1#2{%
 \endgroup
 \patchcmd{\titleblock@produce}
  {\frontmatter@RRAPformat}
  {\frontmatter@RRAPformat{\produce@RRAP{*#1\href{mailto:#2}{#2}}}\frontmatter@RRAPformat}
  {}{}
}%
\makeatother
\begin{document}

\title{A Non-Foster Superconducting Broadband Matching Network}
\author{A. K. Yi}
    \affiliation{SLAC National Accelerator Laboratory, 2575 Sand Hill Road, Menlo Park, California 94025, USA}
    \affiliation{Kavli Institute for Particle Astrophysics and Cosmology, Stanford University, 452 Lomita Mall, Stanford, CA, 94305, USA}
\author{P. Stark}
    \affiliation{Department of Physics, Stanford University, 382 Via Pueblo Mall, Stanford, CA, 94305, USA}
  
\author{C. Bartram}%
    \affiliation{SLAC National Accelerator Laboratory, 2575 Sand Hill Road, Menlo Park, California 94025, USA}
\affiliation{Kavli Institute for Particle Astrophysics and Cosmology,
Stanford University, 452 Lomita Mall, Stanford, CA, 94305, USA}
\date{\today}
\begin{abstract}
The nonlinear inductance of the Josephson junction has enabled the development of a wide range of continuous-variable amplifiers and qubit-based devices with unprecedented sensitivity. We present an alternative use of the Josephson junction in the context of broadband impedance matching. The idea poses a potential solution to a longstanding problem in the field of high energy particle physics. The axion, a compelling candidate for the dark matter, converts to a weak electromagnetic signal at an as-yet unknown frequency. As such, the ideal axion detector does not compromise bandwidth for sensitivity, a trade-off intrinsic to all linear, time-invariant and passive circuits. We propose a circuit that uses a Josephson junction in an impedance matching network to overcome these gain-bandwidth constraints and increase the scan rate of axion searches. The Josephson junction can be biased to exhibit negative inductance capable of canceling geometric inductance similar to a capacitor but across a wider frequency range.
\end{abstract}

\maketitle

\section{Introduction}
Broadband impedance matching is an important area of research within antenna, microwave, and radio frequency (RF) engineering ~\cite{pozar2021microwave,chen2015broadband,collin2007foundations}. Impedance matching is a \textcolor{black}{core} design consideration in network analysis to minimize signal loss due to internal reflection and maximize transfer efficiency between a signal source and load. However, a classical impedance-matched complex network can only achieve optimal matching over a finite bandwidth. 

This limitation is described by the Bode-Fano limits, which constrain the average value of the inverse magnitude of the reflection coefficient as a function of frequency for classical passive, linear, and time invariant (LTI) networks with complex loads~\cite{fano1950theoretical,bode1945network}. This gain-bandwidth product imposes stringent constraints on the design of broadband matching networks, and considerable efforts have been dedicated to developing network architectures that present bandwidth improvements over the classical limit. 

\begin{figure}[t]
    \centering
    \includegraphics[width=0.85\linewidth]{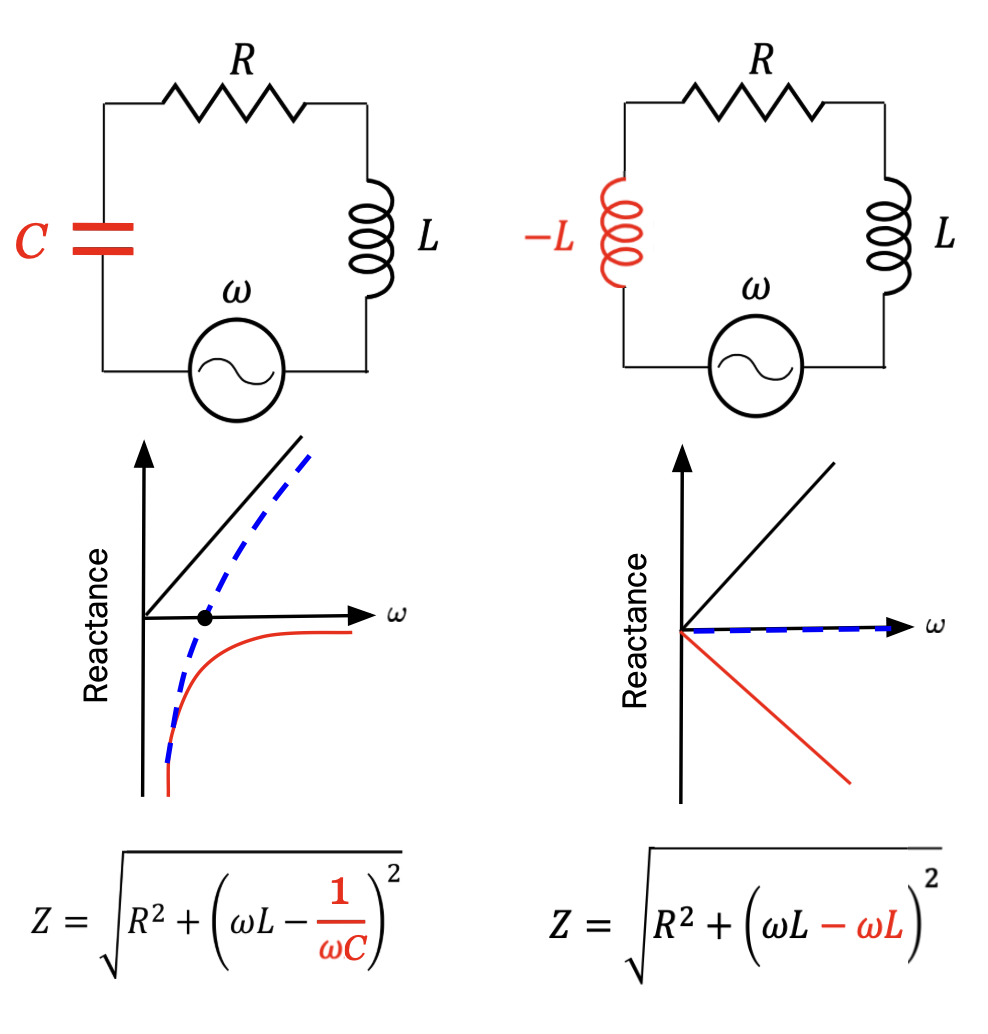}
    \caption{Left: In a classical, single-pole resonant matching network, the reactive impedance is canceled at a single value, indicated by the zero crossing of the total reactive inductance. Right: In a non-Foster network with idealized negative inductance instead of capacitance, the reactive impedance is canceled at all frequencies, resulting in a technically infinite bandwidth network.}
    \label{fig:nonfoster}
\end{figure}
\textcolor{black}{One approach to evading the Bode-Fano limits is} through the use of \textit{non-Foster} networks, which incorporate active devices that violate the Foster reactance theorem and subvert the passive LTI assumptions behind the Bode-Fano limits~\cite{fosterreactance1924}. These networks are often realized as active semiconductor circuits with negative impedances, such as negative impedance converters (NICs) constructed out of operational amplifiers~\cite{hrabar_non-_nodate,merrill1951theory,larky1957negative,yuce2008negative}. Historical approaches have especially focused on high quality factor (high-$Q$) networks within the radio and microwave frequency band. \textcolor{black}{Such} systems tend to be dominated by reactive impedances, rather than radiative losses, which leads to a narrowband network where the impedance is only perfectly matched at a single \textit{resonant} frequency. In contrast, an ideal non-Foster network with a negative reactive impedance could cancel the reactance at \textit{all} frequencies, resulting in an infinite bandwidth device, as shown in Fig.~\ref{fig:nonfoster}. Existing applications of non-Foster networks at these frequencies include electrically small antennas for broadband communication and imaging~\cite{renHighGainCompact2021, jacobBroadbandNonFosterMatching2012, loghmanniaBroadbandParametricImpedance2022b,sussman2009non,mekawy2021parametric}, dispersion compensation in active metamaterial and 2D structures~\cite{Kharina01032000, Hraber201089,hrabar_first_2018}, high power transmission line design~\cite{jacob2017non, long2014stable}, fast-wave propagation~\cite{long2014broadband} and superluminal waveguides~\cite{sievenpiper2011superluminal}. 

A newfound driver in the development of non-Foster networks arises in the field of high energy particle physics. A compelling dark matter candidate known as the axion is predicted to convert to an electromagnetic signal in a magnetic field~\cite{sikivie1983experimental}. The axion of quantum chromodynamics, or QCD axion, is thought to be highly motivated due to its ability to simultaneously explain a related problem in high energy physics~\cite{Peccei1977June,Wilczek:1977pj,Weinberg:1977ma,PhysRevD.96.095001,Dine:1981rt,Zhitnitsky:1980tq,DINE1983137}. Once thought too weak to detect, signals of the requisite strength for the QCD axion were first rendered visible through the combination of a high-$Q$ resonator and SQUID amplifier~\cite{asztalos2010squid,du2018search}. Nevertheless, this circuit's necessarily high quality factor imposes strict limits on its bandwidth in accordance with the Bode-Fano constraints. As such, coverage of the full parameter space proceeds at an excruciatingly slow pace, with estimates of required scan time at tens of thousands of years~\cite{malnou2019squeezed}.

The concept of improved impedance matching to the dark matter has been discussed in general terms, but, to our awareness, no specific implementation along these lines has been proposed. In analogy to the bandwidth limitations on passive LTI matching networks, a rigorous derivation on the fundamental limitations to the scan rate of axion searches has been developed~\cite{chaudhuri2018optimal, chaudhuri2021impedance}. In contrast, in this work, we present a circuit design that is no longer limited by the aforementioned bandwidth constraints. 

One requirement in the search for the axion is the maintenance of the extraordinary sensitivity achieved through the use of cryogenic superconducting materials. We propose a technique that fulfills this requirement while achieving a match over a wider bandwidth than that of a passive, LTI circuit. Our technique leverages the negative inductance of the Josephson junction as part of the impedance matching network that functions as the axion detector.

\section{Circuit Design}
\begin{figure}
    \centering
    \includegraphics[width=0.5\linewidth]{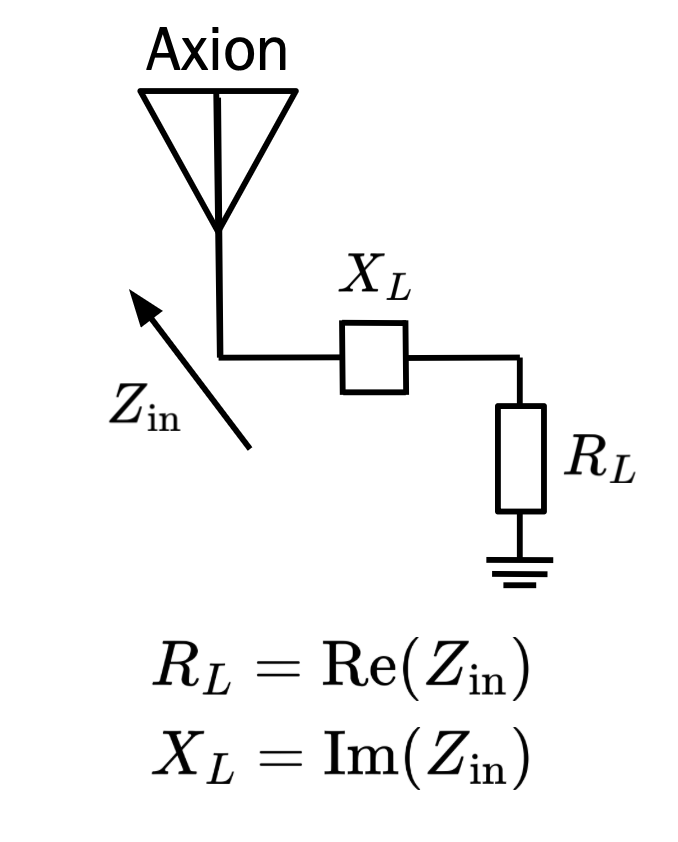}
    \caption{Load receiver model for a signal with a complex source impedance. For a detector to achieve a perfect match, the real impedance of the load must match the real component of the source impedance, while the complex load impedance must equal the complex conjugate of the complex source impedance.}
    \label{fig:load_receiver}
\end{figure}

In this section, we introduce a Josephson junction-based circuit design that achieves negative inductance as part of a non-Foster device. The proposed device is intended to serve as a matching network between a complex-valued source impedance and the signal readout (Fig.~\ref{fig:load_receiver}). 

This model is especially relevant in the context of lumped element axion searches, which operate in a regime where the primary observable for the axion is a dark matter-induced magnetic field. Under these assumptions, the signal arising from the axion dark matter can be modeled as a perfectly stiff AC voltage source. Axion detectors \textcolor{black}{couple inductively to this signal}, which introduces a complex term to the source impedance that must be matched by the network to maximize the transfer of signal power incident on the pickup to the readout. \textcolor{black}{For a predominantly inductive source impedance,} the most efficient narrowband match is a capacitor, creating a resonant network that cancels the inductance of the pickup at a single frequency. \textcolor{black}{Our proposed device instead replaces the capacitor with a Josephson junction device biased to a negative inductance state to act as the non-Foster component in the matching network.}

\textcolor{black}{In Sections \ref{subsec_flux_conserv} and \ref{subsec_extended_circuit},} we derive the analytic expression for the total effective inductance and discuss the choice of circuit parameters needed to achieve effective negative inductance. In subsequent sections, we simulate and compare the power transfer within our matching network to that of the classical resonant match. Finally, we discuss the \textcolor{black}{stability} of the proposed device, as well as its implications for axion searches.

\subsection{Theory of the Josephson junction}\label{subsec_jj}

\textcolor{black} {The element at the center of the proposed design is the Josephson junction, a cryogenically operated active device that consists of two superconductors interrupted by an insulating barrier.}

\textcolor{black}{When the superconducting regions are weakly coupled, a  non-dissipative supercurrent $I$ related to the phase difference $\phi$ between their macroscopic wave functions flows across the junction. Furthermore, if the phase difference changes with time, a voltage $V$ develops across the junction. One can use these relations (known as the Josephson equations\cite{Josephson:1962}) to derive an effective inductance for the Josephson junction:}
\begin{equation}
\label{eq_jj_inductance}
    L_J(\phi)=\frac{\Phi_0}{2{\pi}I_c\cos\phi} = \dfrac{L_{J,0}}{\cos\phi},
\end{equation}
This sinusoidal relationship implies that inductance across the junction takes on negative values for ${\pi}/2 < \phi < 3{\pi}/2$. Previous works have demonstrated this inductance-phase relation and discussed the negative inductance regime, but not in the context of broadband impedance matching \cite{rifkin1976current,martinis2004superconducting}. 

\begin{figure}[t]
\centering
\begin{subfigure}{0.25\textwidth}
\begin{circuitikz} 

\draw (0,2.05) node[ocirc] {};
\draw (0,-0.05) node[ocirc] {};
\draw (0,0) to[josephsonjunction, -] (0,2);

\draw[-Triangle, very thick](0.5, 1) -- (1.15, 1);

\draw (2.5, 2.05) node[ocirc] {};
\draw (2.5,-0.05) node[ocirc] {};
\draw
(2.5,0) to[short, -] (2.5,0.25)
(1.75,0.25) to[short, -] (2.5,0.25)
(1.75,0.25) to[capacitor, -] (1.75,1.75)
(2.5,0.25) to[josephsonjunction, -] (2.5,1.75)
(3.25,0.25) to[resistor, -] (3.25,1.75)
(2.5,0.25) to[short, -] (3.25,0.25)
(2.5,1.75) to[short, -] (3.25,1.75)
(2.5,1.75) to[short, -] (1.75,1.75)
(2.5,1.75) to[short, -] (2.5,2)
;
\end{circuitikz}
\caption{}
\end{subfigure}%
\begin{subfigure}{0.25\textwidth}
\centering
\includegraphics[width=0.9\linewidth]{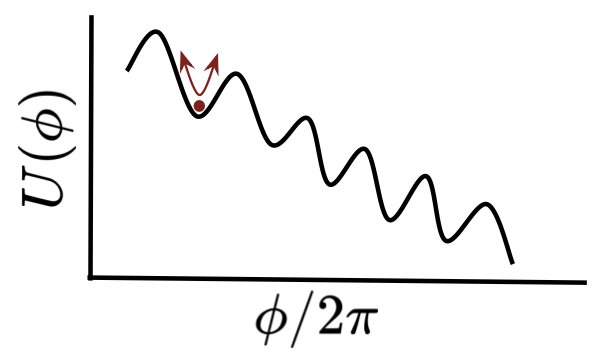}
    \caption{}
\end{subfigure}
\caption{\textcolor{black} {(a) Realistic RCSJ model of a Josephson junction. All simulations employ the RCSJ model with non-zero values for the shunt resistance and capacitance. (b) Washboard potential of a RCSJ junction. Junction resistance and capacitance values introduce inertial terms that modify the dynamical equation-of-motion of the phase.}}
\label{fig:RCSJ}
\end{figure}

\textcolor{black}{In practice, real fabricated junctions are better described by the RCSJ-model shown in Fig.~\ref{fig:RCSJ}. In that case, the dynamical behavior of the junction phase is described by the ``tilted washboard potential". When fabricating the junction, one normally wishes to reduced hysteretic behavior by selecting the resistance and capacitance values of the junction, denoted as $R_J$ and $C_J$ respectively.}

\subsection{Flux quantization in a superconducting loop} \label{subsec_flux_conserv}

\begin{figure}[b]
\includegraphics[width=0.15\textwidth]{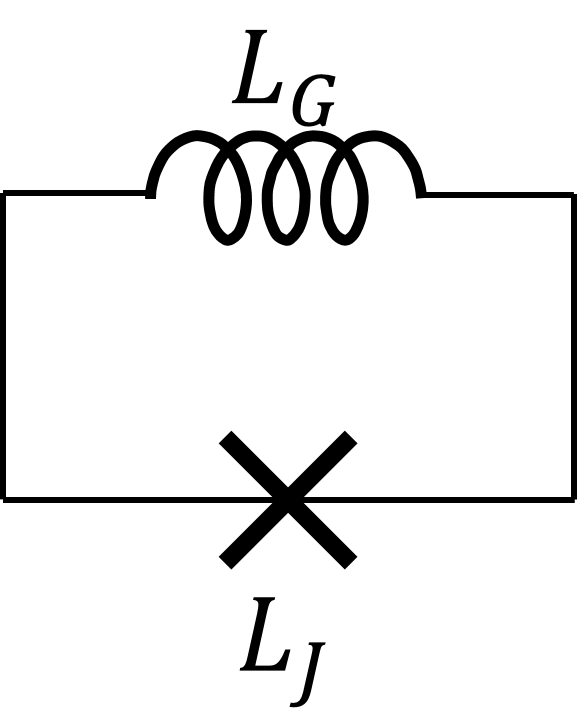}
    \caption{A superconducting circuit loop with a geometric inductance, whose value is denoted by $L_{G}$, and Josephson junction with inductance $L_{J}$.}
    \label{fig_loop}
\end{figure}

To understand the operating principle of a potential non-Foster device, we consider the simple example of a flux-biased superconducting loop interrupted by a Josephson junction (Fig.~\ref{fig_loop}). \textcolor{black} {Any such loop will necessarily possess inductance associated with the geometric inductance of the wire, which we denote as $L_G$.}

If the loop is threaded with an steady state external flux $\Phi_{\mathrm{ext}}$, a current is induced within the loop according to the first Josephson equation. Taking into account the loop inductance, the total flux in the loop is given by
\begin{align}\label{eq:flux_loop}
    \nonumber \Phi &= \Phi_{\mathrm{ext}} + L_GI \\
    \nonumber &= \Phi_{\mathrm{ext}} - L_GI_c\sin(2\pi\Phi/\Phi_0) \\
    &= \Phi_{\mathrm{ext}} - \dfrac{\Phi_0}{2\pi L_{J, 0}} L_G \sin(2\pi\Phi/\Phi_0),
\end{align}
where $I$ is the current induced in the loop and $\Phi_0$ is the magnetic flux quantum. In the last expression, we write the critical current of the junction in terms of the reduced junction inductance, $L_{J,0}$, as defined in Eq.~\ref{eq_jj_inductance}. \textcolor{black}{If the phase difference across the junction is denoted as $\phi$, then the total phase change along the loop is
\begin{equation}
    \delta = \phi + \dfrac{2\pi}{\Phi_0}\Phi.
\end{equation}
}

\textcolor{black}{Because the wavefunction on the loop must be continuous, the phase difference $\delta$ must be an integer multiple of $2\pi$, such that the phase across the junction equals
\begin{equation}\label{eq:phase_diff}
    \phi = 2\pi n - \dfrac{2\pi}{\Phi_0}\Phi.
\end{equation}}

\textcolor{black}{Finally, we isolate $\Phi$ in Eq.~\ref{eq:phase_diff} and substitute it back into Eq.~\ref{eq:flux_loop} to cast the flux loop expression in terms of the junction phase, $\phi$, as}
\begin{equation} \label{eq_loop}
    \dfrac{\Phi_0}{2\pi}\left[\phi - \phi_0 + \dfrac{L_G}{L_{J,0}}\sin\phi\right] = n\Phi_0  ,
\end{equation}
where $n$ is an integer imposing flux quantization and $\phi_{0}$ is the external phase bias associated with external applied flux $\Phi_{\mathrm{ext}} = (\Phi_{0}/2\pi)\phi_{0}$.

The ratio of $L_G / L_{J, 0}$ is the \textcolor{black}{screening} parameter, $\beta_L$, which determines whether the above expression is single-valued (non-hysteretic) or multi-valued. \textcolor{black}{In the non-hysteretic mode, the junction loop behaves as a parametric inductor, with a one-to-one relationship between the phase across the junction and the external applied flux, $\Phi_{\text{ext}}$.}

\textcolor{black}{Though we can use any junction phase ${\pi}/2 < \phi < 3{\pi}/2$ that yields a negative inductance, maximally negative inductance occurs at $\phi = \pi$, which we will use as the optimal phase bias point for impedance matching for the rest of this discussion.} Note that this choice of optimal bias point only holds for non-hysteretic operation. \textcolor{black}{For hysteretic operation, Eq.~\ref{eq_loop} becomes multi-valued and admits multiple meta-stable solutions for initial phase $\phi_0 = \pi$. Simulations run in this mode demonstrate that the $\phi=\pi$ phase solution is highly unstable and preferentially slips to bias points that result in a positive inductance, making it difficult to prepare the initial negative inductance state.}  

\subsection{Extended Circuit Diagram} \label{subsec_extended_circuit}
\begin{figure}
\centering
\includegraphics[width=0.3\textwidth]{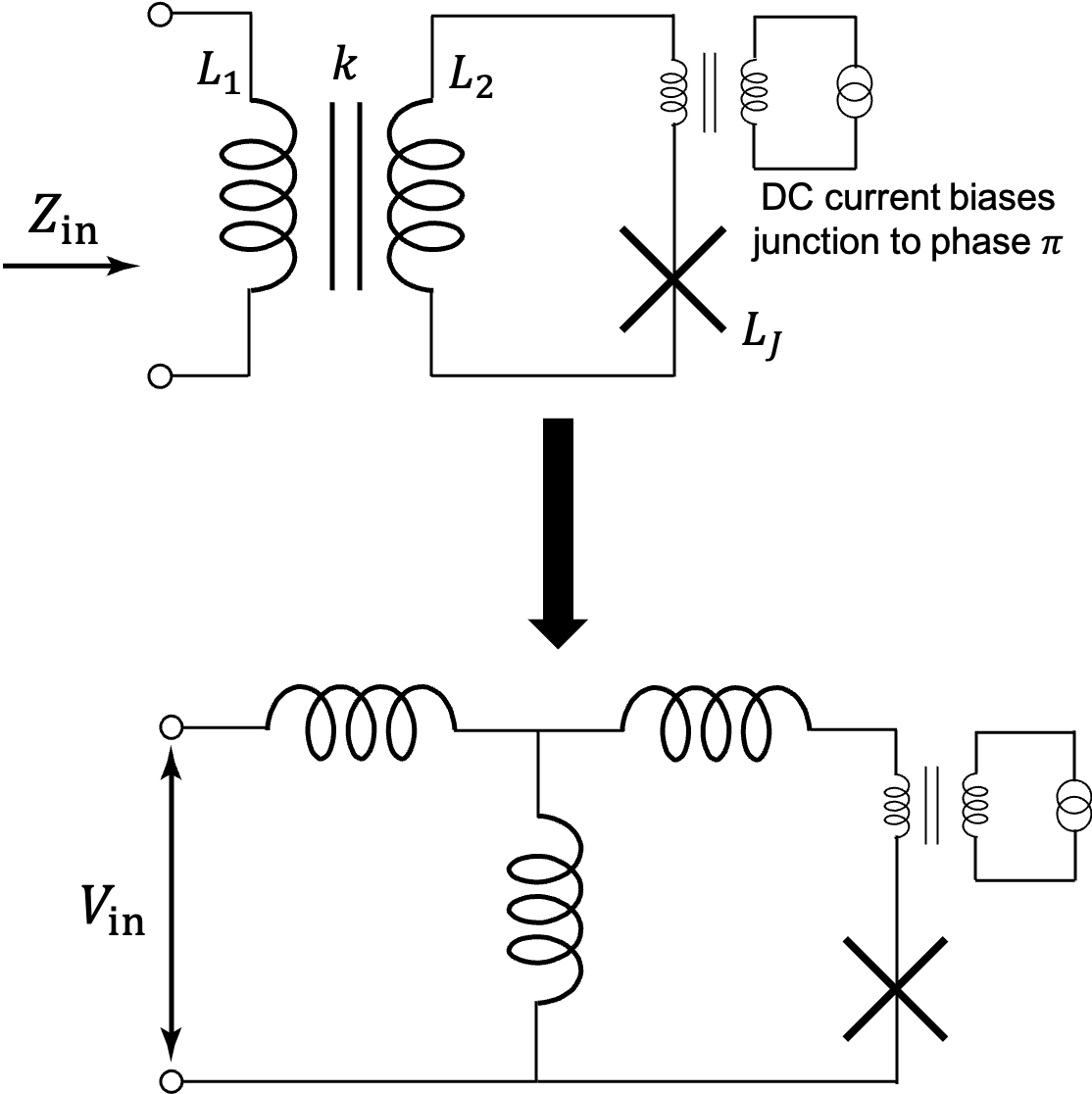}
    \caption{Proposed non-Foster junction device with its equivalent circuit representation. The Thevenin equivalent look-back impedance at that port is $Z_\mathrm{in}$, with a corresponding voltage $V_\mathrm{in}$ across the terminals. The inductor that weakly couples in the magnetic flux from the DC current bias has an inductance much smaller than other sources of inductance.}
    \label{fig_expanded_bfe}
\end{figure}

\textcolor{black}{In Sec.~\ref{subsec_flux_conserv}, we showed that the junction can be reliably biased to negative inductance as long as the geometric inductance seen by the junction is smaller than the reduced junction inductance, $L_{J,0}$. In realistic applications, the junction will ``see" many other sources of inductance on its input port aside from the geometric self-inductance of the loop. However, we can generalize the constraint found in Sec.~\ref{subsec_flux_conserv} to show that for the junction to remain non-hysteretic, the equivalent total inductance seen by the junction must be smaller than $L_{J,0}$.} The proposed device (pictured in Fig.~\ref{fig_expanded_bfe}) expands on this principle to incorporate the junction into a generalized AC matching network. The device is now coupled to a broader network and its input port now includes its inductive coupling to the rest of the detector, as well as the upstream inductive pickup of our axion or other radio-frequency signal detector.

We initially bias the junction to phase $\pi$ using an inductively coupled current bias to prepare the negative inductance state as described in Sec.~\ref{subsec_flux_conserv}. \textcolor{black}{When an RF signal appears on the input of the device, an oscillatory component is introduced to the current within the junction loop and the external applied flux defined in Eq.~\ref{eq:flux_loop} gains a time-dependent component $\Phi_{\text{ext}} = \Phi_{\pi} + \Phi_{rf}$, where $\Phi_{\pi}$ is the original DC current that yields $\phi=\pi$ and $\Phi_{rf}$ is the alternating perturbation on top of it.}

\textcolor{black}{When the driving frequency of the junction is much smaller than both the junction characteristic frequency, $\omega_c = 2\pi I_c R_J/\Phi_0$, and the junction plasma frequency $\omega_{p} = 1/\sqrt{L_JC_J}$, we are operating in what is known as the adiabatic regime. In this regime, the external flux changes slowly enough that the junction phase can always settle in a local minimum of the tilted washboard potential. For sufficiently small signals, where the induced currents across the junction are much smaller than its critical current, the phase across the junction $\phi(t)$ gains a small time-dependence, but remains biased to an average phase very close to $\pi$. In this limit, the junction acquires the effective inductance introduced in Eq.~\ref{eq_jj_inductance}.}

For an arbitrary coupling coefficient $k$, the effective inductance of the non-Foster device is
\begin{equation}
\begin{split}
    L_{\text{eff}} &= L_{1} \left( 1 - \frac{k^{2}L_{2}}{L_{2}+L_{J}}\right)
    \\ &= L_{1} \left( 1 - \frac{k^{2}L_{2}}{L_{2}+L_{J,0}/\cos\phi}\right)
    .
\end{split}
\label{eq_l1_effective}
\end{equation}

To ensure that the phase remains biased to the expected negative inductance value, we consider what the Josephson junction views as its ``geometric inductance." Generalizing the total matching network as series and parallel inductances across the input nodes of the junction yields the following expression:
\begin{equation}
    L_{G} = L_{2} \left( 1 - \frac{k^{2}L_{1}}{L_{1}+L_{0}}\right)
    .
\label{eq_lg_effective}
\end{equation}

Here, $L_{0}$ is any upstream inductance that is not directly present within the non-Foster device loop. In our case, this is the \textcolor{black}{pickup inductance} associated with the coupling of our signal source to our network, which is the source of impedance we want to match to achieve broadband power transfer.

\textcolor{black}{In accordance with the discussion in Sec.~\ref{subsec_flux_conserv}, when $L_{G} < L_{J,0}$, applying an external phase bias equal to $\phi_0 = \pi$ (or, a flux bias equal to $\Phi_0/2$) will result in a junction phase $\phi = \pi$, with a negative junction inductance $L_{J} = -L_{J,0}$.} If $L_{2} + L_{J} > 0$, $L_{\mathrm{eff}}$ will be negative for certain values of $k$ and $L_{2}$, especially if $k$ is sufficiently large or $L_{J,0} \lesssim L_{2}$. Eq.~\ref{eq_lg_effective} will be a consideration throughout the subsequent section, where $L_{G}$ will be split into multiple inductors in configurations where they are in series, in parallel, or mutually inductive. Despite the expansion of the circuit, \textcolor{black}{we control the final phase of the Josephson junction such that the effective inductance across the device is negative.}

\textcolor{black}{There are several important caveats to the above discussion. First, we do not consider any parasitic resistance or capacitance in the network when calculating $L_{G}$. These parasitics will change the junction model, which may lead to incorrect selection of fabrication values or unexpected dynamical behavior that results in runaway phase behavior in the junction. This can present challenges in maintaining a stable matching network, which will be further considered in Sec.~\ref{subsec_stability}. However, it is important to note that Eq.~\ref{eq_lg_effective} is predominantly used for the initial selection of junction values. The resistance and capacitance are incorporated in simulation and used to fine-tune the chosen device parameters by performing a parameter search around the analytical guess.}

\textcolor{black}{Finally, the discussion in the above section focuses primarily on the conditions necessary to prepare the junction loop at a negative inductance. In reality, this device is operated as part of a broader matching network with sources of resistive and inductive impedance (shown in Fig.~\ref{fig_bfe}). Therefore, stable operation of the device depends not only on whether we can bias it to $\pi$, but whether it is stable over time within the broader network. We impose stability by requiring that the total inductance of the network is positive. Further discussion on the stability of the network can be found in Sec.~\ref{subsec_stability}.}

\section{Simulation} \label{sec_simulation}
From the results of Sec.~\ref{subsec_extended_circuit}, we construct a circuit that uses the non-Foster component for impedance matching and simulate this using WRSPICE~\cite{WRSPICE}. WRSPICE is a circuit simulation tool based on LTSPICE which supports a variety of Josephson junction models. In this work, we use a WRSPICE model based on the RSCJ model~\cite{WRSPICE1982rsjmodel}, \textcolor{black}{with a capacitance of 0.7 nF per critical current in amperes and a subgap resistance of 16.5 k$\Omega$. For the chosen critical current value of 1 \si{\micro\ampere}, the junction plasma frequency is roughly 330 GHz.}

\begin{figure}
\centering
\includegraphics[width=0.45\textwidth]{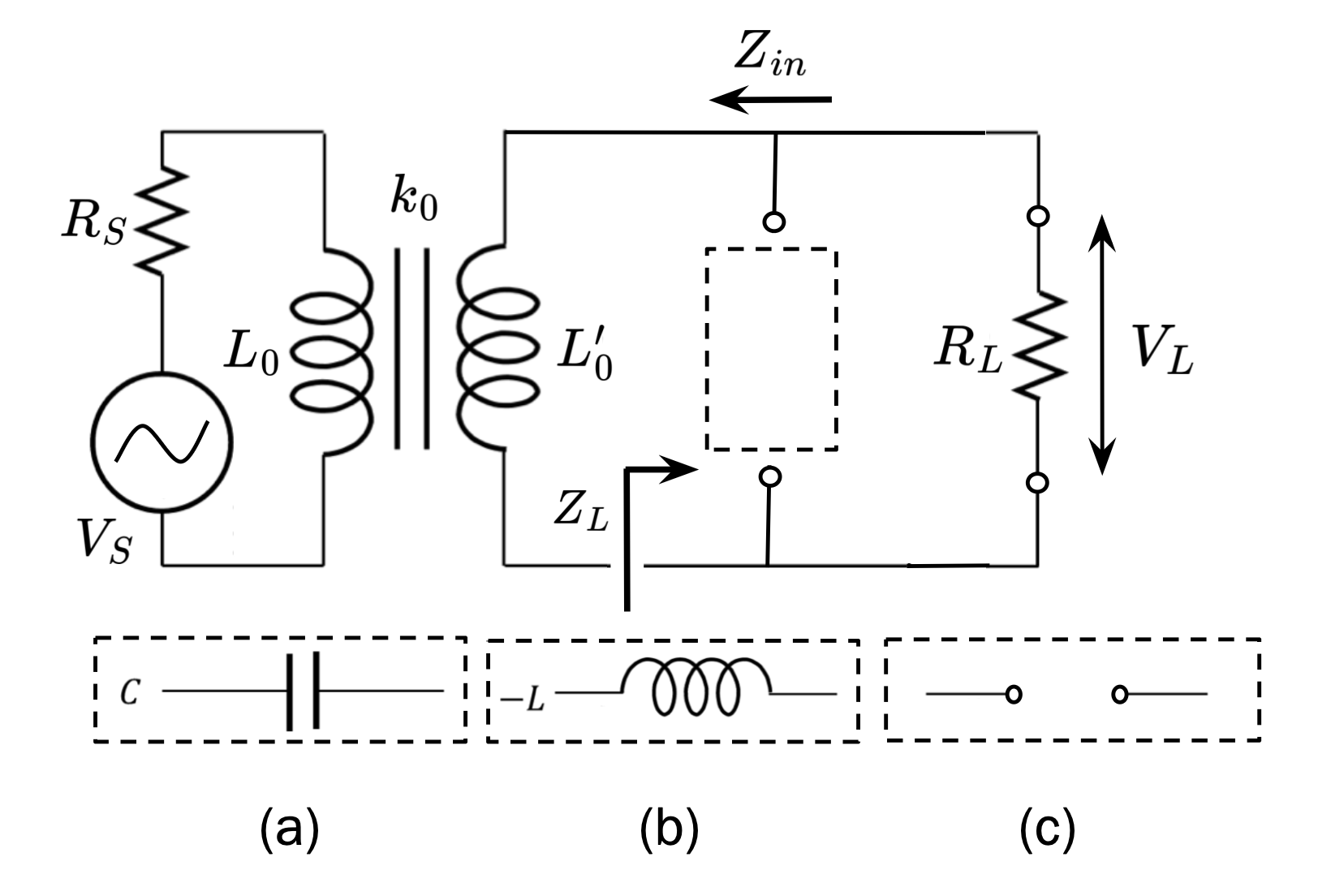}
    \caption{Full signal model and matching network. Simulations are performed to compare three different impedance matching networks in a $\pi$-network configuration: (a) classical capacitor match, (b) negative inductance match with the device proposed in Fig.~\ref{fig_expanded_bfe}, and (c) no match. The output voltage and subsequent power is measured across a 50 $\Omega$ load resistor $R_L$.}
    \label{fig_bfe}
\end{figure}

Simulated voltages, currents, and phases are obtained in the time domain. Using this data, the average input and output power are calculated, and their ratio is taken as the value for power transfer. We define input power as the real power supplied by the voltage source and the output power as the power dissipated across the 50 $\Omega$ load resistor. To calculate average power in the time domain, we take the instantaneous product of the current and voltage, integrate over multiple wavelengths, and then divide by the total time interval. This is equivalent to calculating real power using the product of root-mean-square voltage and current and the phase angle between them.

\textcolor{black}{To see the power transfer over a band of voltage driving frequencies, we performed a frequency sweep} from 30 MHz to 70 MHz with a step size of 1 MHz. Individual simulations were performed for each frequency, and final voltage and current values were measured at timescales past the initial transient behavior of the circuit. We compare the wideband power transfer of the non-Foster circuit to an optimal capacitive match, as well as to a circuit with no matching component at all (see Fig.~\ref{fig_bfe}). In terms of axion experiments, the former are analogous to resonant experiments such as DMRadio and ADMX, and the latter to broadband experiments like ABRACADABRA~\cite{alshirawi2025electromagnetic,rosenberg1998search,salemi2021search}.

\subsection{Power Transfer Comparison Between Impedance Matching Circuits} \label{subsec_power_transfer}

\begin{figure*}
    \centering
    \begin{minipage}{.48\textwidth} 
        \centering
        \includegraphics[width=\linewidth]{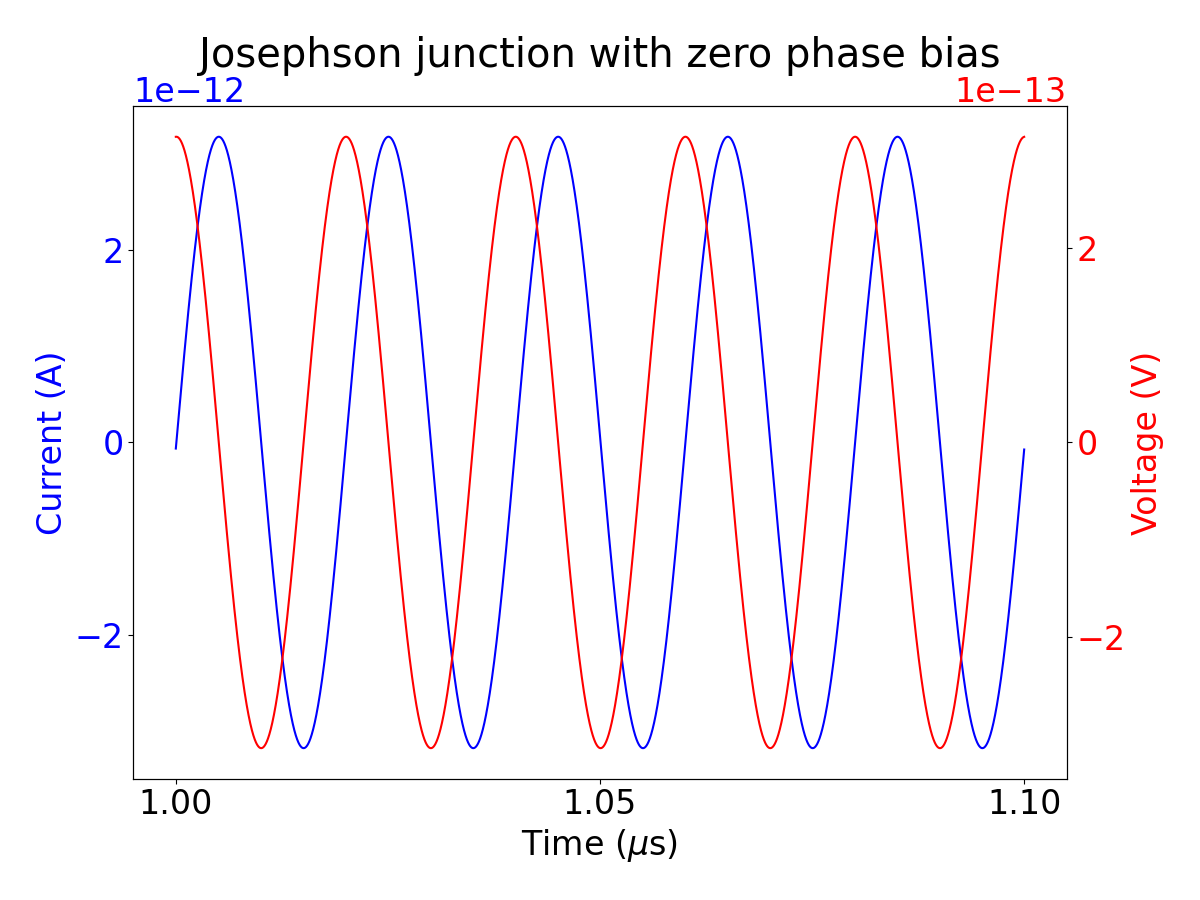} 
    \end{minipage}
    \hfill 
    \begin{minipage}{.48\textwidth} 
        \centering
        \includegraphics[width=\linewidth]{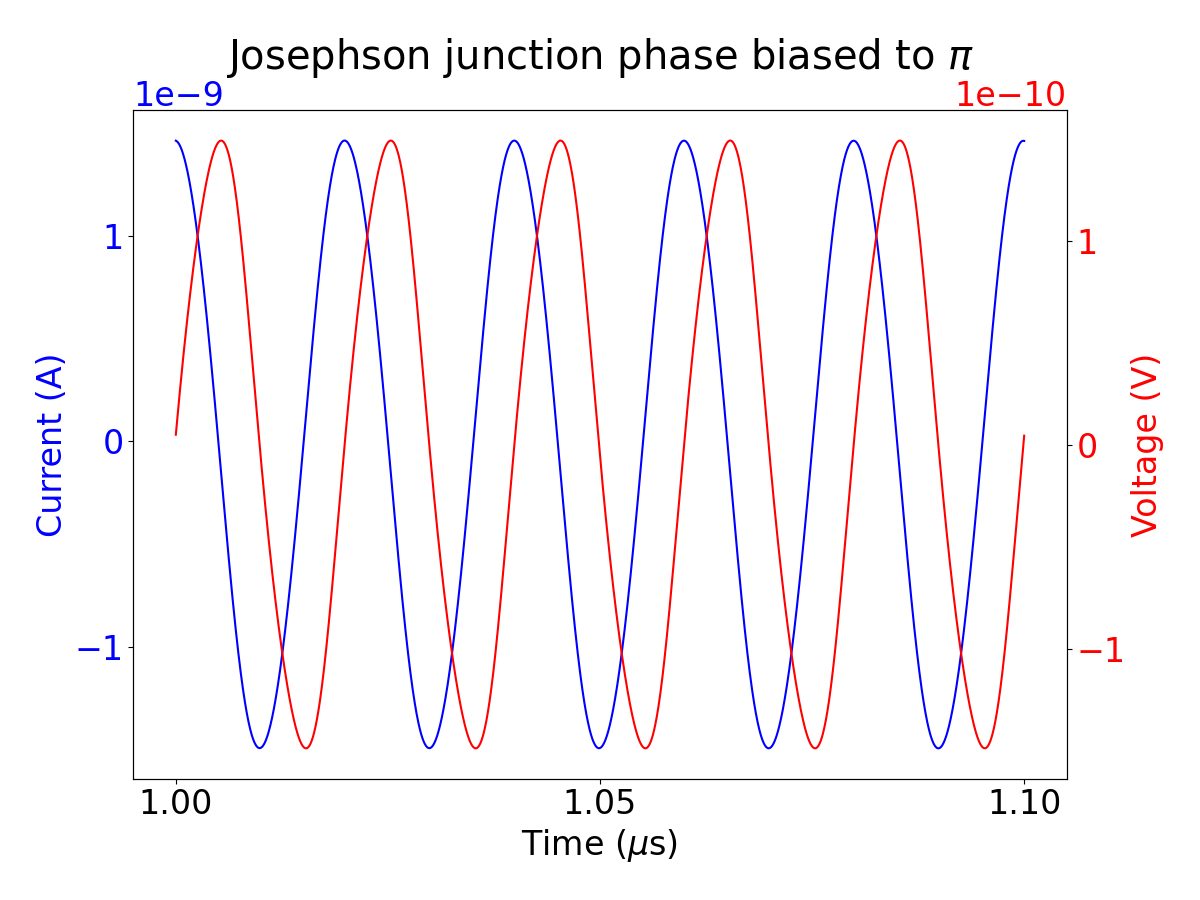}
    \end{minipage}
    \caption{Time domain simulation results in the non-Foster matching circuit when the Josephson junction's phase is initially biased to zero (left) and $\pi$ (right). The current (blue) and voltage (red) across $L_{1}$ are shown for each simulation. In the case where there is zero phase bias, the current lags behind the voltage, behaving like a normal inductor. This indicates that $L_\mathrm{eff}$ is positive. On the other hand, when the junction is phase biased to $\pi$, the current leads the voltage. In this case, $L_\mathrm{eff}$ is negative, and the phase of the voltage has flipped 180 degrees compared to the above simulation. As the effective inductance cancels out the rest of the inductance in the circuit, we can achieve an impedance match. This is also evident as the amplitudes of both current and voltage are higher in the latter plot.
    }
    \label{fig_phase_bias_jj_comp}
\end{figure*}

The resonant circuit with a capacitor in parallel will be used as a passive, linear, and time invariant baseline for evaluating power transfer. In designing this circuit, the input and output resistances were chosen to be 5 $\Omega$ and 50 $\Omega$, respectively, with a corresponding quality factor $Q = 500$. The resonant frequency of the circuit was chosen to be 50 MHz, fixing the values for $L_{0} = L_{0}^{\prime} = 3.183\times10^{-10}$ H, the coupling constant $k_{0}=0.316$, and the capacitor $C = 3.183\times10^{-8}$ H.

While the input inductance is split between two mutually inductive inductors ($L_{0}$ and $L_{0}^{\prime}$), the relatively large value of $R_{S}$ make the effective input inductance equal to $L_{0} = L_{0}^{\prime}$, which keeps the use of Eq.~\ref{eq_lg_effective} consistent. This assumption is valid up to voltage source frequencies in the GHz range, where other considerations such as parasitic capacitance begin to invalidate the current circuit model.

The inductances were deliberately chosen to have $\mathcal{O}(10^{-10})$ H values in consideration of typical Josephson junction inductances at critical currents around 1 \si{\micro\ampere}. Circuits with higher $Q$ values could have also been chosen \textcolor{black}{and would more realistically model the smaller input resistances associated with superconducting circuits.} In this work, we opted for lower values of $Q$ with higher input resistance to reduce the time needed for simulations to reach steady state behavior.

The second matching circuit uses the non-Foster matching network introduced in Sec.~\ref{subsec_extended_circuit} (Fig.~\ref{fig_expanded_bfe}). The effective contribution of the non-Foster component must have the same impedance as the capacitor in the resonant circuit. There are many combinations of $L_{1}$, $L_{2}$, and $k$ that satisfy Eq.~\ref{eq_l1_effective} and produce a negative $L_{\mathrm{eff}}$ which gives the required impedance matching of the circuit. In this work, we use a combination with $L_{1}=3.297\times10^{-10}$ H, $L_{2}=3.449\times10^{-10}$ H, and $k=0.3$. With a higher $k$, better power transfer can be achieved, but 0.3 was selected as a reasonable value that could be achieved during circuit fabrication. $L_{1}$ and $L_{2}$ were designed to be similar values and optimized to produce the highest power transfer. The Josephson junction is set to have a critical current of 1 \si{\micro\ampere}, which corresponds to an inductance of $3.291\times10^{-10}$ H when biased to zero phase. Assuming the Josephson junction settles at phase $\pi$, the effective inductance $L_\mathrm{eff}$ is equal to $-3.183\times10^{-10}$ H. \textcolor{black}{Though the initial phase can be set directly through WRSPICE, we instead couple in magnetic flux through a small inductor inserted into the junction loop to mimic the inductive coupling of} a bias line on a fabricated chip. Its inductance is $10^{-14}$ H and has minimal effect on the total inductance of the circuit.

\textcolor{black}{To confirm that the junction exhibits negative inductance, we compare two simulations of the current and voltage across the input of the device: (1) when the junction phase has zero bias (positive inductance) and (2) when the junction phase is $\pi$ (negative inductance).} Both simulations use a 50 MHz AC voltage source, with a voltage amplitude of $10^{-10}$ V. \textcolor{black}{The results of the simulations are shown in Fig.~\ref{fig_phase_bias_jj_comp} and demonstrate that biasing the junction to $\pi$ results in the current of the device leading its voltage instead of lagging it by $90^\circ$, as would be typical for an inductor.} In other words, $L_\mathrm{eff}$ obeys the expected voltage-current relation of an inductor, while exhibiting an additional $180^\circ$ phase shift that generates the effective negative inductance. This provides an impedance match similar to the capacitor, but does so for a wider range of frequencies, since the frequency dependence of the source inductance and $L_\mathrm{eff}$ is identical.

Finally, the circuit is simulated without any matching component. In comparison to the resonant circuit, this circuit fully removes the parallel matching branch of the capacitor. The transfer of power is fully dependent on inductive reactance that has not been canceled, and therefore we expect its efficiency to suffer greatly. 

\begin{figure}
\centering
\includegraphics[width=0.5\textwidth]{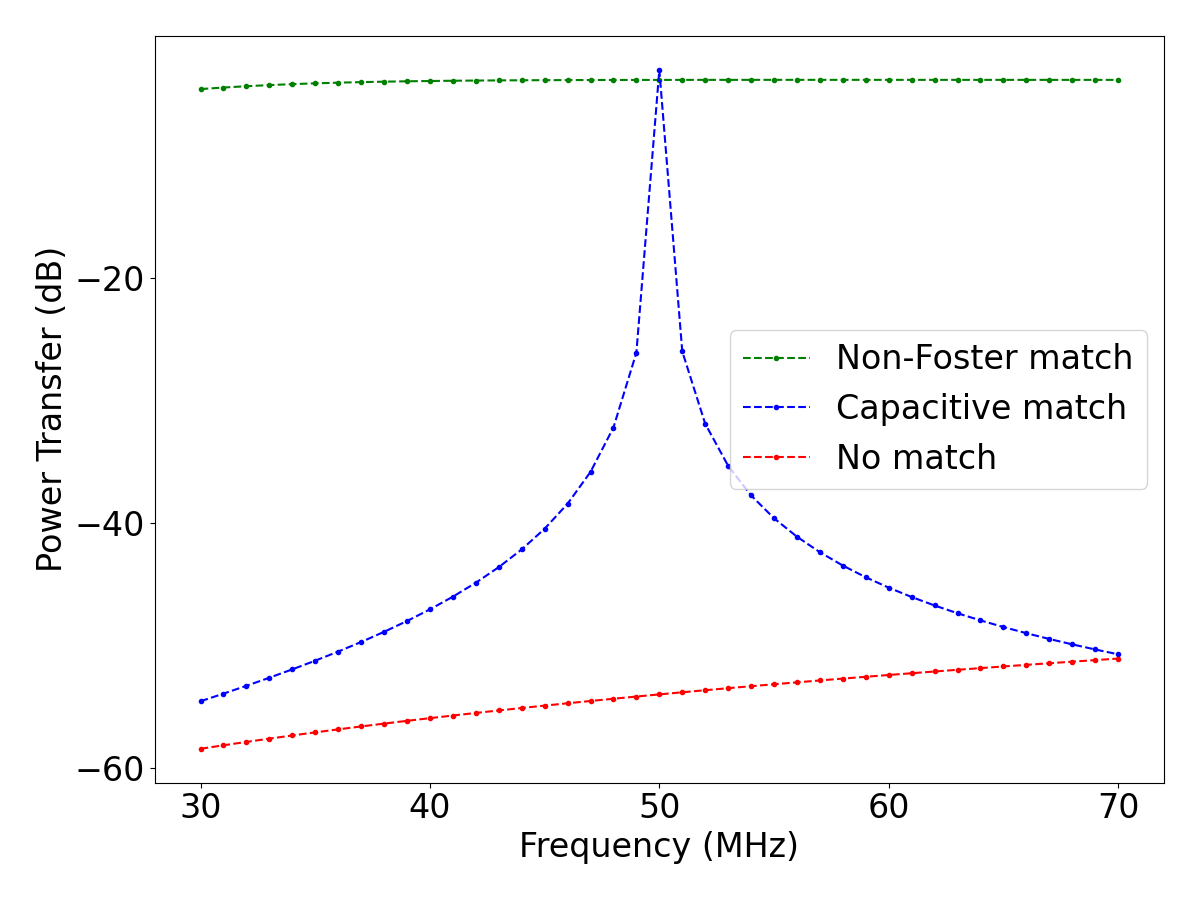}
    \caption{The simulated power transfer comparison between a capacitively-matched resonant circuit (blue), a non-Foster impedance match circuit (green), and a circuit without any matching at all (red), from 30 MHz to 70 MHz. The resonance frequency of the resonant circuit is 50 MHz. Each point is a separate WRSPICE simulation done in the time domain, and power is calculated at a time scale after any initial, transient behavior in the circuit.
    }
    \label{fig_power_transfer_comp}
\end{figure}

\begin{figure*}
    \centering
    \begin{minipage}{.48\textwidth} 
        \centering
        \includegraphics[width=\linewidth]{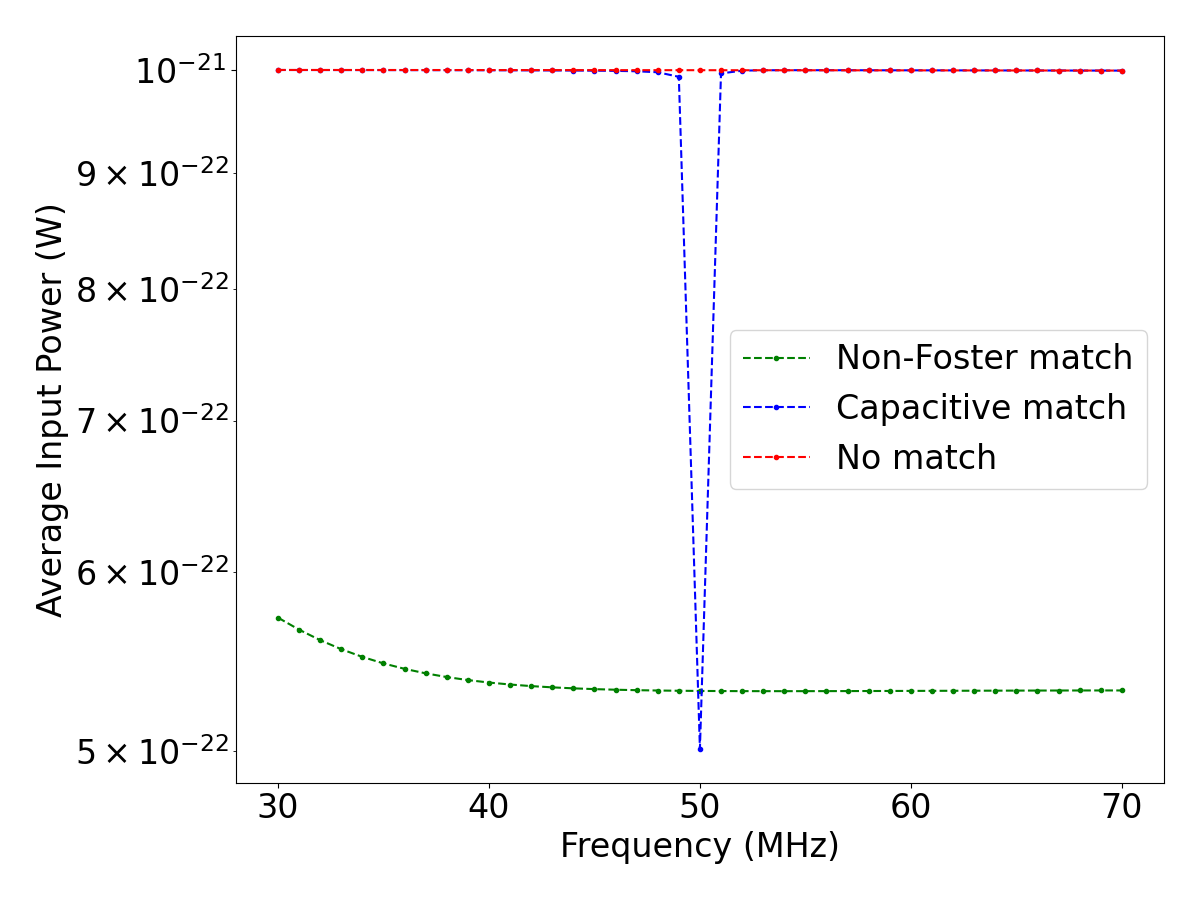} 
    \end{minipage}
    \hfill 
    \begin{minipage}{.48\textwidth} 
        \centering
        \includegraphics[width=\linewidth]{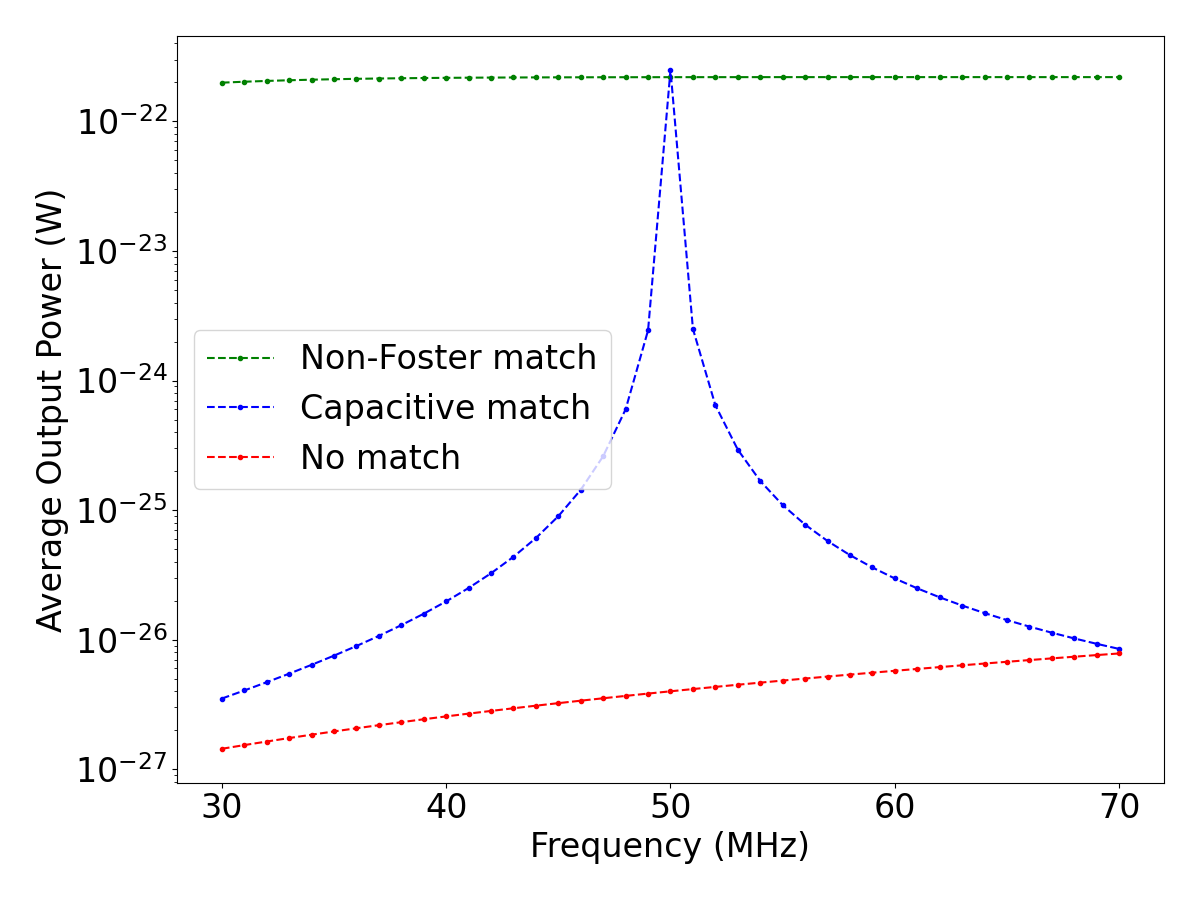}
    \end{minipage}
    \caption{Average input power (left) and output power (right) comparison between a capacitively-matched resonant circuit (blue), a non-Foster impedance match circuit (green), and a circuit without any matching at all (red), from 30 MHz to 70 MHz. The resonance frequency of the resonant circuit is 50 MHz. Each point is a separate WRSPICE simulation done in the time domain, and power is calculated at a time scale after any initial, transient behavior in the circuit. The input power is calculated at the voltage source, and the output power is calculated at the output resistor.} 
     \label{fig_power_input_output_comp}
\end{figure*}

Fig.~\ref{fig_power_transfer_comp} compares the simulated power transfer between all three circuits. For the capacitively matched network (shown in blue), the power \textcolor{black}{transfer is characteristic of a typical resonant circuit with $Q=500$. At resonance, the input and output resistors are optimally matched and exactly half the input power is delivered to the load, resulting in a 3 dB loss at 50 MHz.} The power transfer of the non-Foster network is shown in the green line in Fig.~\ref{fig_power_transfer_comp}. Similarly to the resonant circuit, the average power is calculated after the initial transient behavior of the Josephson junction. In particular, we consider the power transfer at the \si{\micro\second} scale, from 1 \si{\micro\second} to 2 \si{\micro\second}. The \textcolor{black}{long-term} stability of this circuit (i.e., change in power transfer as a function of time) is discussed in the following subsection. On these timescales, the power transfer is shown to be about -4 dB across the frequency range of interest, as shown in the green dotted line of Fig.~\ref{fig_power_transfer_comp}. Overall, this shows that a broadband matching circuit can be achieved with the given non-Foster configuration with only a 1 dB difference from optimal transfer. The power transfer with no matching network is seen in red on Fig.~\ref{fig_power_transfer_comp}. Since there is no impedance matching, the efficiency in power transfer is low. The non-Foster circuit power transfer is better by about 50 dB.

As seen in the simulation results above, not only is the power transfer of the non-Foster circuit similar to that of a resonant impedance matching circuit, but it also exhibits an extended bandwidth. One consideration for the circuit is the power input and output of each type of circuit that was compared, shown in the top plot of Fig.~\ref{fig_power_input_output_comp}. Due to the design of the circuit, the average power drawn by the voltage source decreases by roughly 50\% at resonance for the resonant circuit. This decrease in power drawn by the voltage source can also be seen for the wider bandwidth of the non-Foster circuit, with values similar but slightly higher to the resonant circuit at resonance. Nevertheless, due to the high rates of power transfer, the total power that is dissipated by the output resistor is naturally higher when the impedance is matched (right plot of Fig.~\ref{fig_power_input_output_comp}). While the resonant circuit does this for a small bandwidth of frequencies around 50 MHz, the non-Foster matching circuit achieves similar results with only about 13\% decrease in output power from 30 MHz to 70 MHz.

\begin{figure}[t]
\centering
\includegraphics[width=\columnwidth]{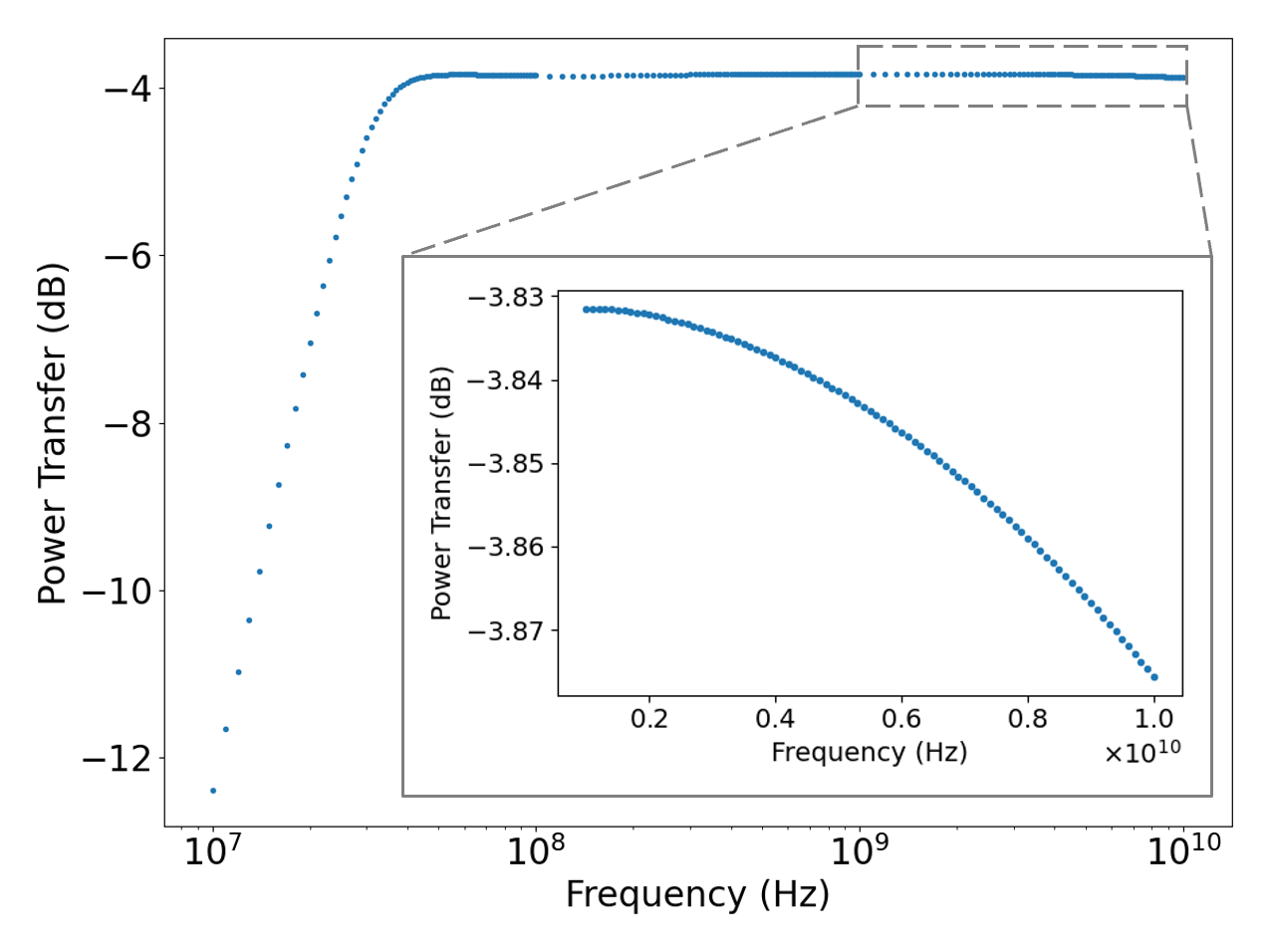}
    \caption{The simulated power transfer of the previously described non-Foster impedance matching circuit. Simulations were done from 10 MHz to 10 GHz\textcolor{black}{, with the inset showing the power transfer above 1 GHz, where a small drop in power transfer is observed}. Each point is a separate WRSPICE simulation done in the time domain, and power is calculated at a time scale after any initial, transient behavior in the circuit.
    }
    \label{fig_power_wideband}
\end{figure}

Following the promising results from the non-Foster impedance matching circuit across 30 MHz to 70 MHz, an additional simulation was done to determine the full bandwidth for which the circuit maintains its impedance matching capabilities. In Fig.~\ref{fig_power_wideband}, the non-Foster circuit was simulated with WRSPICE for multiple orders of magnitude, from 10 MHz to 10 GHz. Simulations with higher voltage source frequencies were not performed since the circuit model \textcolor{black}{begins to break down at} frequencies on the order of GHz. At these frequencies, \textcolor{black}{we begin to depart from the lumped element regime and} additional considerations such as stray capacitance would be required for a more realistic simulation. \textcolor{black}{Input voltage source frequencies at hundreds of GHz will also be around the Josephson junction's plasma frequency, and therefore were avoided in the wideband simulations.} Nevertheless, for the current set of simulations, the power transfer of the circuit is better than -4 dB from 30 MHz and onwards. This suggests that it is possible to maintain well-impedance-matched power transfer over a much larger range of frequencies, provided that the circuit model used for the simulation remains valid.

\textcolor{black}{It is important to emphasize that all the simulated circuits were optimized for transfer at 50 MHz, such that the non-Foster matching circuit has the same impedance $L_\mathrm{eff}$ as the capacitor in a 50 MHz resonant circuit. Using the idealized inductance-only equations in Sec.~\ref{subsec_extended_circuit}, the device is expected to achieve an optimal power transfer of -3 dB with a wide bandwidth centered at 50 MHz. However, when the resistor and capacitor of the RCSJ model are more accurately included in the calculations, the maximum power transfer decreases to -3.8 dB due to the additional sources of impedance that cannot be fully matched.}

\textcolor{black}{Finally, because the simulations incorporate the more physically accurate junction model, we do observe frequency dependence in the wideband simulations depicted in Fig.~\ref{fig_power_wideband}. Below 30 MHz, we see a steep drop that corresponds to the point where the total inductance of the circuit falls below zero for the chosen parameters at 50 MHz. This observation is confirmed with simulation, where we see the phase settle away from the optimal value in the stable regions, decreasing power transfer below 30 MHz. At higher frequencies, the drop in efficiency (see inset of Fig.~\ref{fig_power_wideband}) corresponds to the frequencies where the inductive and resistive impedance become similar in magnitude. We note again that at those frequencies, we no longer trust that our model is representative of the practical circuit. Operation at higher frequencies would require a more complete treatment of the relevant dispersive effects.}

\subsection{Stability of the Non-Foster Circuit} \label{subsec_stability}

\textcolor{black} {There are three major considerations when discussing the practicality of the proposed non-Foster matching circuit:}
\begin{enumerate}
    \item \textcolor{black} {the inherent feasibility of the device}
    \item \textcolor{black} {the circuit's sensitivity to fluctuations that may lead to instability over time}
    \item \textcolor{black} {the tolerances that critical parameters must exhibit to achieve the intended negative inductance}
\end{enumerate}

\textcolor{black}{Point (1) is best addressed through a physical argument. It is widely acknowledged that an ideal, dispersionless non-Foster device is impractical because an infinite bandwidth device with ``inverse" impedance would violate basic principles of energy conservation and causality. Indeed, if one could design a stable, infinite bandwidth negative impedance device, one would theoretically be able to source infinite power. This paradox is resolved by realizing that all physically realistic non-Foster devices must inevitably be dispersive. Extensive work has been done to develop a realistic causal model that can explain their behavior and stability conditions \cite{okornPhysicallySoundModel2017,ugarte-munoz_stability_2012}. Particularly relevant are the results presented in Ref.~45, in which the authors evaluate the stability requirements of networks with realistic dispersive negative inductors and demonstrate that a parallel combination of positive and negative inductors is stable for the condition $L < |L_{\text{negative}}|$.}

\textcolor{black}{These considerations are relevant to the non-Foster implementation proposed in this paper. First, the realistic RCSJ junction model used in simulation does incorporate dispersion by including a shunt capacitor and resistor. Though their effects are subdominant in the frequency band of interest, they impose a natural bandwidth limitation on the model developed in this paper. In particular, this device is operated in the adiabatic regime, where the source frequency is much lower than both the plasma and characteristic junction frequencies. As we approach the 330 GHz plasma frequency, we expect our analytical junction model to break down and the results claimed in this paper to no longer apply.}

\textcolor{black}{In addition, though our circuit model is not simply a parallel $L-L_{\text{negative}}$, we can cast our mutually inductive input into a T-circuit with both series and parallel inductance branches. As the input coupling coefficient $k$ approaches 1, the equivalent impedance approaches a purely parallel inductor, allowing us to generalize the above stability conditions to our own device. As discussed in Sec.~\ref{subsec_extended_circuit}, the network must obey the requirements that the equivalent inductance seen by the junction $L_G$ is less than $L_{J,0}$ and the total inductance seen by the voltage source is greater than zero. The former requirement is necessary for non-hysteretic operation, while the latter was imposed on purely physical grounds, since total negative inductance across the voltage source would indeed source infinite power and violate energy conservation. We can see now that this intuition is a generalization of the more rigorous stability requirements for a parallel $L-L_{\text{negative}}$ network~\cite{ugarte-munoz_stability_2012}.}

\textcolor{black}{Point (2) and (3) are related and concern our ability to prepare and maintain optimal matching values over time. We perform three sets of simulations to probe this question. First, we run the simulation for longer and see if the power transfer degrades over time. Second, we increase the error tolerance on the bias current to test the precision needed for optimal operation. Third, we change the configuration of inductor values to show the difference in power transfer when $L_{\text{eff}}$ does not perfectly match its target value. These simulations give hints to the performance of the system under different suboptimal conditions.}

\begin{figure}
\centering
\includegraphics[width=0.5\textwidth]{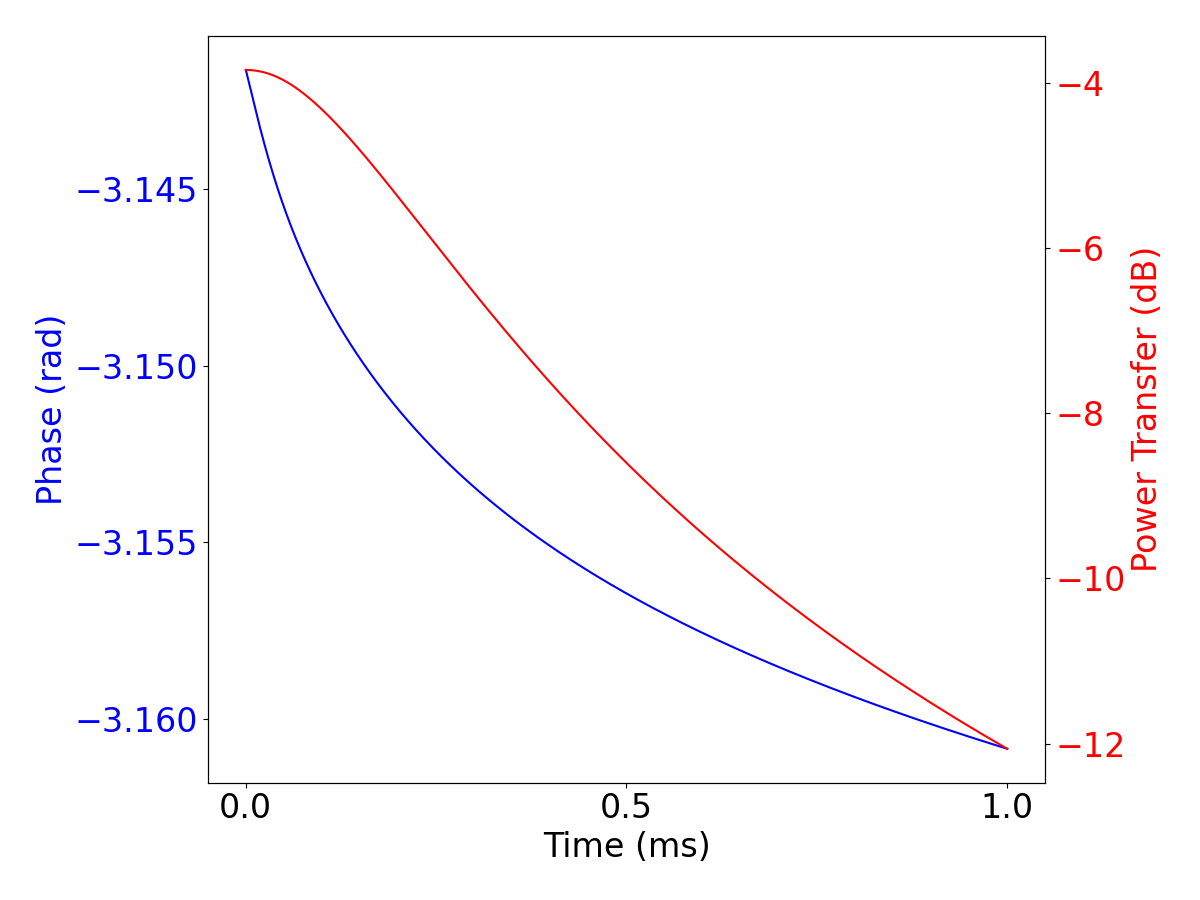}
    \caption{The change in Josephson junction phase and power transfer as a function of time for the non-Foster impedance matching circuit with parameters used in Sec.~\ref{subsec_power_transfer}. The input voltage source has a frequency of 50 MHz, and the simulation was done until 1 ms, as opposed to the \si{\micro\second} scale simulation in the previous subsection. The initial behavior of the Josephson junction is removed for clarity.}
    \label{fig_phase_stability}
\end{figure}

Shown in Fig.~\ref{fig_phase_stability} is the change in power transfer of the non-Foster impedance matching circuit with parameters used in Sec.~\ref{subsec_power_transfer}. As time passes, the Josephson junction phase deviates further away from the initially intended phase $\pi$. After 1 ms, the power transfer drops to -12 dB from the -4 dB at 1 \si{\micro\second}. \textcolor{black}{The source of this instability is observed to relate to the effects of dissipative components in the matching network.} This puts a constraint on the time duration of non-Foster circuit operation.

\begin{figure}
\centering
\includegraphics[width=0.5\textwidth]{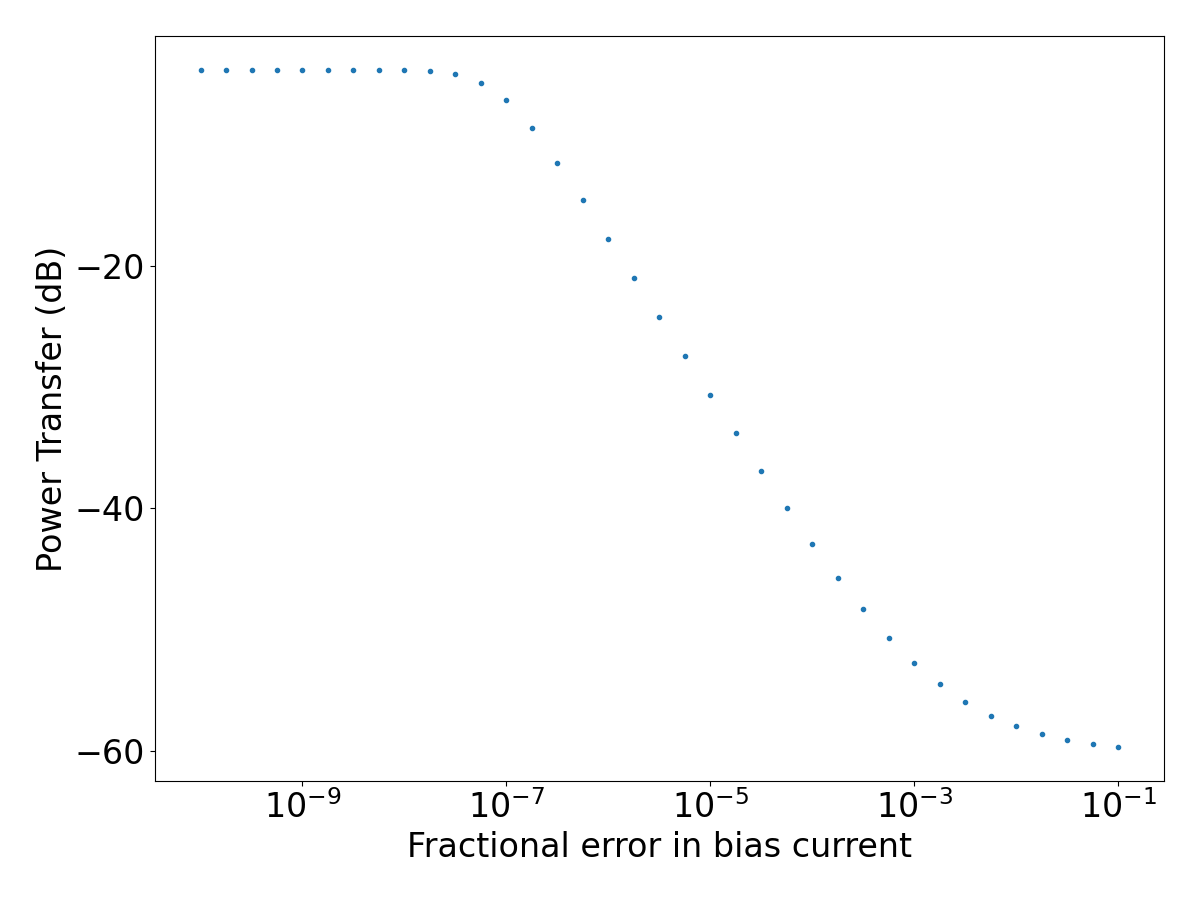}
    \caption{Power transfer as a function of the error in bias current from its optimal point. Power transfer values were taken from time scales of 1 \si{\micro\second}. The power transfer begins to drop from the optimum value as the fractional error approaches $10^{-7}$.}
    \label{fig_bias_current_stability}
\end{figure}

The second simulation, shown in Fig.~\ref{fig_bias_current_stability}, shows how the power transfer changes when there is error in the bias current. The fractional error in the bias current is proportional to the error in the initial phase bias for the Josephson junction, which is intended to be $\pi$. This simulation shows that the required accuracy of the bias current must be less than 100 parts per billion to maintain optimal power transfer. This result puts a constraint on the precision of the bias current, which could be affected by noise in our current source.

\begin{table*}[]
\centering
\renewcommand{\arraystretch}{1.35} 
\setlength{\tabcolsep}{8pt}       

\begin{tabular}{|>{\centering\arraybackslash}c
                |>{\centering\arraybackslash}c
                |>{\centering\arraybackslash}c
                |>{\centering\arraybackslash}c
                |>{\centering\arraybackslash}c
                |>{\centering\arraybackslash}c
                |>{\centering\arraybackslash}c|}
\hline
Configuration & $L_{1}$ & $L_{2}$ & $k$ & A & B & C \\
\hline
$L_{\mathrm{eff}} = -L_{0}$ & 3.297 & 3.449 & 0.3 & -3.844  & -12.061 & -30.625 \\
\hline
$L_{\mathrm{eff}} = -1.005L_{0}$ & 3.747 & 3.459 & 0.3 & -8.035  & -10.451     & -30.154 \\
\hline
$L_{\mathrm{eff}} = -1.01L_{0}$  & 3.498 & 3.453 & 0.3 & -13.331 & -13.797     & -29.615 \\
\hline
\end{tabular}

\caption{Power transfer simulation results for three configurations of the non-Foster impedance matching circuit. The $L_{1}$ and $L_{2}$ values, whose inductances are in units of $10^{-10}$ H, are varied. Based on the values of $L_{1}$, $L_{2}$, and $k$, each configuration has an effective inductance $L_{\mathrm{eff}}$ equal to $-L_{0}$, $-1.005L_{0}$, and $-1.01L_{0}$, respectively. For each configuration, three power transfer results, all in units of dB, are considered. The configurations are as follows. A: power transfer at 1 \si{\micro\second}, B: power transfer at 1 ms, C: power transfer at 1 \si{\micro\second}, with $10^{-5}$ fractional error in bias current. For all configurations, the drive frequency of the voltage source is 50 MHz. }
\label{table_transfer_comp}
\end{table*}

Finally, we discuss the changes in power transfer when the inductor values on the non-Foster component change. This reflects fabrication conditions when the targeted inductance parameters are not perfectly produced. In Table~\ref{table_transfer_comp}, we simulate the non-Foster impedance matching circuit with similar values of $L_{1}$ and $L_{2}$. The change in inductances will have $L_{\mathrm{eff}}$ that does not fully match the reactive inductance in the rest of the circuit. Due to this, the power transfer is lower. There is, however, an advantage in power transfer at longer scales as the dropoff in phase is slower, demonstrating improved stability. This observation suggests that there is a tradeoff between maximum power transfer and the duration over which we can maintain it. Power transfer loss due to bias current imprecision, on the other hand, remains relatively similar when the fractional error is $10^{-5}$. Due to this, maintaining stable bias currents with low fluctuations is expected to be an important task in practice.

\subsection{Maintaining operational negative inductance with alternating pulse currents}

\textcolor{black}{The issues of the long-term decay in Josephson junction phase, and the ultimately power transfer of the non-Foster matching circuit, could be addressed} with a \textit{time-variant} implementation of the bias current.

\begin{figure*}
    \centering
    \begin{minipage}{.48\textwidth} 
        \centering
        \includegraphics[width=\linewidth]{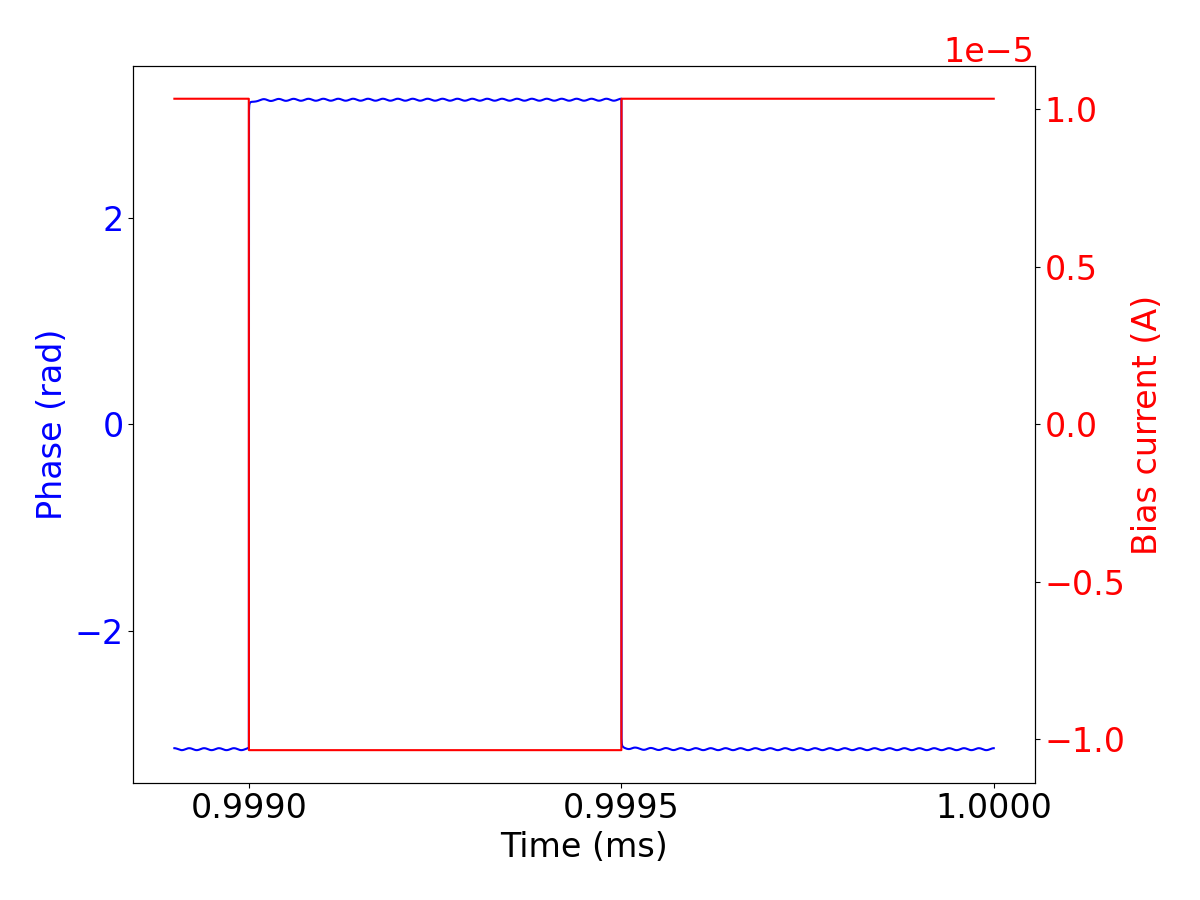} 
    \end{minipage}
    \hfill 
    \begin{minipage}{.48\textwidth} 
        \centering
        \includegraphics[width=\linewidth]{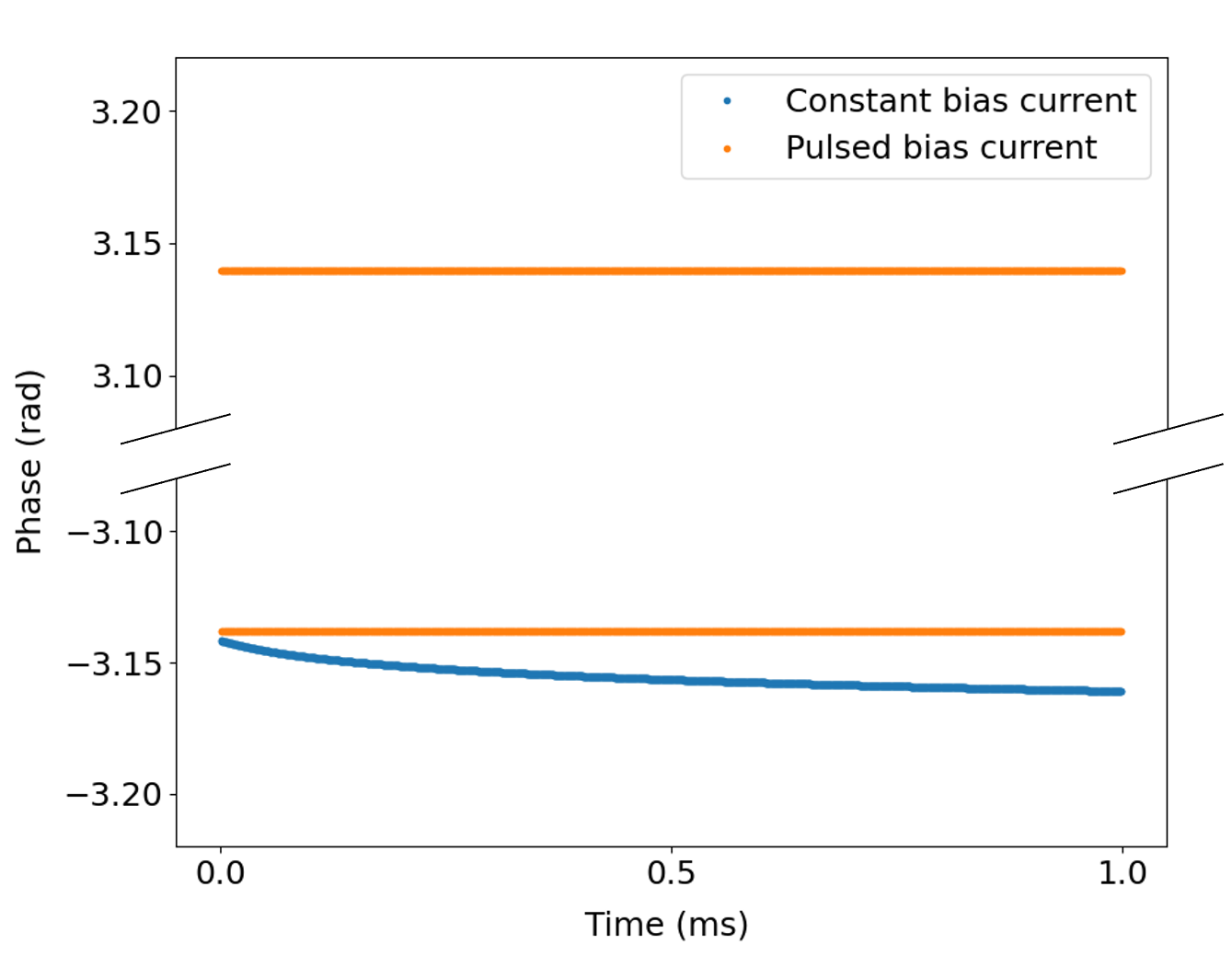}
    \end{minipage}
    \\
    \begin{minipage}{\textwidth} 
        \centering
        \includegraphics[width=\textwidth]{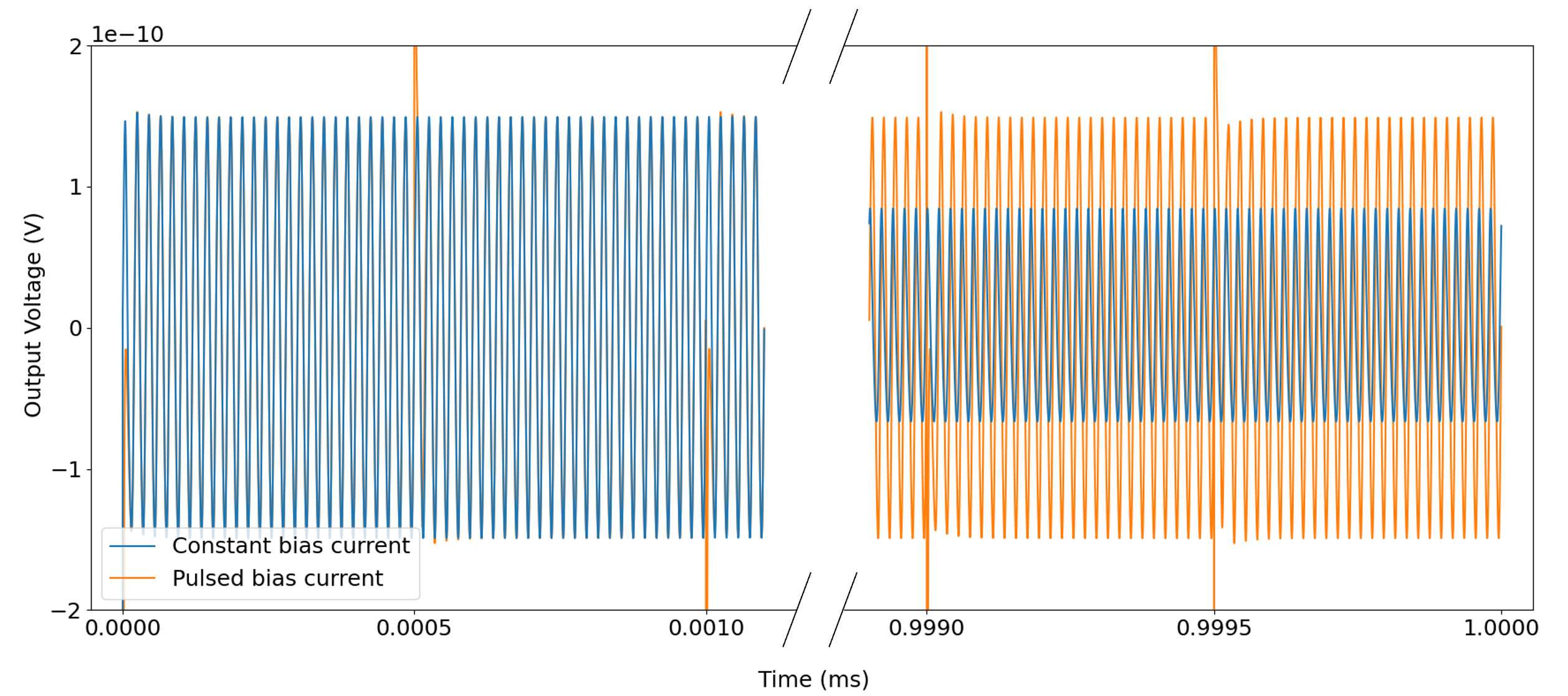}
    \end{minipage}
    \caption{Top left: A pulsed bias current and corresponding Josephson junction phase for a time domain WRSPICE simulation of the non-Foster impedance matching circuit. All parameters are the same as those used in the optimal power transfer simulation, except for the bias current, which changes sign every every 0.5 $\si{\micro\ampere}$. Only the last part of the simulation is shown. The voltage source frequency is 50 MHz, so each change in the bias current sign happens every 25 wavelengths. Top right: A comparison of the phase of the Josephson junction is shown for both constant and pulsed. Phase is averaged across every 0.5 $\si{\micro\ampere}$, equal to the duration of which the pulsed bias current remains at one value. The Josephson junction's phase under a pulsed bias current is either around $\pi$ or $-\pi$, without showing any long term change in phase. The prevention of phase decay is not possible in the constant bias current simulation. Bottom: The output voltage for constant and pulsed bias current simulations. The beginning of the simulation and the end of the simulation are shown. The spikes in the pulsed bias current simulation correspond to each ``pulse" where the bias current changes sign.}
    \label{fig_bfe_pulsing}
\end{figure*}

\textcolor{black}{Before the Josephson junction's phase slips into the regime where the power transfer is suboptimal, its phase} can be re-established through the use of a periodic current pulse with the same magnitude but in the opposite direction. By alternating the sign of the bias current, the junction phase alternates between $\pi$ and $-\pi$. \textcolor{black}{Because the Josephson junction's inductance is an even function of phase, the inductance will remain the same even when the junction is biased to the opposite phase.}

Fig.~\ref{fig_bfe_pulsing} shows the results of a WRSPICE simulation using such periodic pulsing of the bias current. The bias current changes sign every 0.5 $\si{\micro\second}$, or every 25 wavelengths for the voltage source frequency (50 MHz). \textcolor{black}{The choice of pulsing period is somewhat arbitrary;} it is also possible to change the pulse duration to maintain a longer, coherent waveform before the loss in power transfer degrades to a significant level. This optimization will depend on the needs of each experiment that uses the non-Foster impedance matching circuit with a pulsed bias current.

\textcolor{black}{In contrast to a constant bias current, the phase of the junction remains at $\pi$ (or $-\pi$) up to the 1 ms scale when the bias current is pulsed. The output voltage for both simulations, shown in the bottom plot, also suggests that a pulsed bias current keeps the level of output steady up to at least 1 ms, excluding sudden spikes in voltage when the bias current changes sign. This behavior is attributed to the junction being biased to a different phase and requires a short time for it to settle based on values of $R_J$ and $C_J$. Ultimately, the settling time is short compared to the remaining duration of the pulse. Moreover, as evidenced in the beginning of the simulation, the waveforms of the voltage are in phase for both bias current configurations. This means that the change in bias current sign does not affect the coherence of the voltage waveform. By removing the data where there are spikes in voltage, it would be possible to sustain high power transfer for longer periods of time. This would also preserve the knowledge of phase information as it is unaffected outside of the voltage spikes.}

\section{Application to Axion Searches} 
\label{axion_search}

The rate of coverage of the axion parameter space, also known as the scan rate, depends on the model of the detector. Currently, axion searches below a few hundred MHz use lumped-element circuit models, similar to what has been described in prior sections. Above a few hundred MHz, it becomes difficult to use a lumped-element circuit, as stray inductance in the circuit will lower the resonant frequency. This barrier necessitates a different detector model at higher frequencies, motivating a transition to microwave cavities. An interesting aspect of the proposed circuit design is that inductance cancellation could remove or shift the existing delineations between detector models within the axion frequency space.

Among cavity searches, a separate issue further impedes the scan rate. The simplest resonator that will fit in a solenoidal magnet bore is a right cylindrical cavity, whose volume, $V$, scales with frequency, $\nu$, as ${\nu}^{-3}$. Accounting for the frequency scaling of the quality factor ($Q\sim\nu^{-2/3}$), the scan rate  scales as ${\nu}^{-14/3}$. This precipitous decline in the scan rate at higher frequencies is one of the largest challenges in the field, and compounds the issue of the Bode-Fano limits on passive matching networks that exists across all frequencies~\cite{malnou2019squeezed}. The technique proposed here may be applicable in this regime, but a different circuit model is needed to fully evaluate the the efficacy of the concept at frequencies above a GHz.

A complete understanding of the possible axion scan rate cannot be made without a full accounting of the noise in any detector. Through the Dicke-Radiometer equation, the total system noise, $T_{\mathrm{sys}}$, and the integration time, $t$, are related to the signal-to-noise (SNR) through $\mathrm{SNR}\,{\propto}\,\sqrt{t}/{\mathrm{T_{sys}}}$. Active components, and therefore non-Foster circuits, necessarily contribute more to the total system noise. For biased Josephson junctions operating in nonlinear regimes, the utility of a quantitative noise model is limited without an experimental realization. A few possible noise mechanism include critical current fluctuations, quasiparticle-related noise, bias current noise coupling into junction phase fluctuations, and conversion of phase noise into amplitude noise when operating near $\phi \approx \pi$. A complete understanding of the noise requires a dedicated measurement of the signal-to-noise improvement with and without the non-Foster component. 

While a precise accounting of the noise is not possible without a prototype, we can make a very rough approximation of the impact on the SNR. In a typical axion search, the detector consists of the resonator read out by a low noise quantum amplifier. Since our device leverages the Josephson junction, which is a fundamental building block of these quantum amplifiers, one could assume there is added noise in the resonator at the level of a typical quantum amplifier, such as a SQUID. Since the total system noise is additive, this essentially doubles the amplifier noise associated with existing axion detectors.  All else held equal, including the instantaneous bandwidth of the detector, if it takes time $\tau$ to achieve a given SNR with a passive capacitive matching circuit, then, assuming a doubling of the noise, a non-Foster circuit will take time 4$\tau$ to achieve the same sensitivity. 

Our results suggest a possible increase in the instantaneous bandwidth by a factor of $\sim$1,000. If we assume a doubling of the amplifier noise (and a factor of four increase to the required integration time), the scan rate would then increase by a factor of $\sim$250. Furthermore, the proposed technique eliminates the need for slow and complex mechanical tuning mechanisms that frequently jam and, through frictional heating, raise the base temperature of the cryostat. Nevertheless, a thorough study of the added noise in a variety of implementations and real lab conditions, is critical to assessing the benefit to axion searches.

An interesting aspect of this approach is that the broadband capabilities of the readout necessarily entail faster signal processing. Data acquisition could become a bottleneck in the absence of thorough consideration. Fortunately, digital signal processing has made great strides in recent decades. At high frequencies, this endeavor could stand to benefit from high-speed, high-throughput Radio Frequency System on a Chip (RFSoC) systems with sampling rates as high as 10 GSPS. The telecommunications industry has driven the advancement of these systems to allow for rapid gains in the speed of digital signal processing. The consolidation of various controls and data acquisition instrumentation onto a single RFSoC would provide the ability to tightly synchronize the feedback controls necessary for junction rebiasing with the data-taking operation. We propose that, through the use of an automated rebiasing scheme that provides a fixed duty cycle, axion data could be acquired in the intervals between pulses and co-added via standard analysis procedures offline. Operations under these conditions would require a robust understanding of the repeatability and rate of phase roll-off for the duration of an experimental measurement.

\section{Conclusion and Future Directions}

In conclusion, we present a conceptual design for a Josephson junction operated as a non-Foster circuit component. We believe this circuit represents one possible solution to the longstanding scan rate problem facing axion searches, with a potential increased to the instantaneous bandwidth by a factor of $\sim$1,000. To further evaluate the efficacy of this technique, we propose to fabricate the non-Foster circuit on a chip and demonstrate the power sourced and transferred from the AC input as shown here. Of particular importance is a robust characterization of the circuit noise and stability. Mitigation of instability, through the application of a controlled pulse signal, may be one step towards the implementation of this technique in an axion experiment. We note that, though challenging, the stability issue may represent a more readily solvable problem than those currently impeding the scan rate of existing searches. Finally, low-noise, broadband detection of electromagnetic signals is relevant to a variety of fields, such as astronomy, nuclear, and condensed matter physics, especially in the context of spectroscopy~\cite{formaggio2012project,ji2025local,aybas2021quantum}. As such, we believe this approach may find other applications beyond the more immediate implementation in the detection of axion dark matter.

\textcolor{black}{This work is a proof-of-concept for a simulated superconducting non-Foster impedance matching circuit. Based on the findings of this work, we plan to fabricate a device which closely follows the parameters laid out in the Sec.~\ref{sec_simulation}. Measurements of such a physical device will also give us proper characterization of the noise, such as the added noise from the non-Foster component of the circuit.}

\section{Acknowledgements}
The authors are extremely grateful to the DMRadio and ADMX collaborations for thoughtful conversation and feedback on the manuscript. They also extend thanks to David Schuster and Noah Kurinsky for their insightful commentary on quantum techniques and device physics.

Chelsea Bartram, Andrew Yi and the conceptual idea, design and simulations presented in this work were supported by the Department of Energy, Laboratory Directed Research and Development program at SLAC National Accelerator Laboratory, under contract DE-AC02-76SF00515 and as part of the Panofsky Fellowship awarded to Chelsea Bartram. Pamela Stark is supported by the National Science Foundation Graduate Research Fellowship under Grant No. DGE-2146755.

\section*{Author Declarations}
The authors have no conflicts to disclose.

\section*{Data Availability Statement}
The data that support the findings of this study are available from the corresponding author upon reasonable request.

\bibliography{bodefano-clean}

\end{document}